\documentclass[preprint2]{aastex701}

\begin{document}

\title{The AURORA Survey: Determining the Production Mechanism for O {\sc i} $\mathbf{\lambda 8449}$ Emission in Star-forming Galaxies at Cosmic Noon}

\author[0000-0003-1249-6392]{Leonardo Clarke}
\affiliation{University of California, Los Angeles, 475 Portola Plaza, Los Angeles, CA, 90095, USA}
\email{leoclarke@astro.ucla.edu}

\author[0000-0003-3509-4855]{Alice E. Shapley}
\affiliation{University of California, Los Angeles, 475 Portola Plaza, Los Angeles, CA, 90095, USA}
\email{aes@astro.ucla.edu}

\author[0000-0002-4585-3985]{Zhiwei Shao}
\affiliation{Department of Astronomy, School of Physics and Astronomy,
Shanghai Jiao Tong University, Shanghai 200240, China}
\affiliation{State Key Laboratory of Dark Matter Physics, \& Tsung-Dao Lee Institute, Shanghai Jiao Tong University, Shanghai 200240, China}
\email{}

\author[0000-0002-2282-8795]{Massimo Pascale}
\affiliation{University of California, Los Angeles, 475 Portola Plaza, Los Angeles, CA, 90095, USA}
\email{}

\author[0000-0003-4792-9119]{Ryan L. Sanders}
\affiliation{University of Kentucky, 505 Rose Street, Lexington, KY 40506, USA}
\email{}

\author[0000-0001-9687-4973]{Naveen A. Reddy}
\affiliation{Department of Physics \& Astronomy, University of California, Riverside, 900 University Avenue, Riverside, CA 92521, USA}
\email{}

\author[0000-0001-5860-3419]{Tucker Jones}
\affiliation{Department of Physics and Astronomy, University of California, Davis, 1 Shields Avenue, Davis, CA 95616, USA}
\email{}

\author[0000-0003-1561-3814]{Harley Katz}
\affiliation{Department of Astronomy \& Astrophysics, University of Chicago, 5640 S Ellis Avenue, Chicago, IL 60637, USA}
\affiliation{Kavli Institute for Cosmological Physics, University of Chicago, Chicago IL 60637, USA}
\email{}

\author[0000-0001-9489-3791]{Natalie Lam}
\altaffiliation{NSF Graduate Research Fellow}
\affiliation{University of California, Los Angeles, 475 Portola Plaza, Los Angeles, CA, 90095, USA}
\email{}

\author[0009-0002-6186-0293]{Shreya Karthikeyan}
\affiliation{University of California, Los Angeles, 475 Portola Plaza, Los Angeles, CA, 90095, USA}
\email{}

\author[0000-0002-7622-0208]{Callum T. Donnan}
\affiliation{NSF’s National Optical-Infrared Astronomy Research Laboratory, 950 N. Cherry Ave., Tucson, AZ 85719, USA}
\email{}

\author[0000-0003-4264-3381]
{N. M. F\"orster Schreiber}
\affiliation{Max-Planck-Institut f\"ur extraterrestrische Physik (MPE), Giessenbachstr.1, D-85748 Garching, Germany}
\email{}

\author[0000-0003-4464-4505]{Anthony J. Pahl}
\affiliation{The Observatories of the Carnegie Institution for Science, 813 Santa Barbara Street, Pasadena, CA 91101, USA}
\email{}

\author[0000-0002-4153-053X]{Danielle A. Berg}\affiliation{Department of Astronomy, The University of Texas at Austin, 2515 Speedway, Stop C1400, Austin, TX 78712, USA}
\email{}






\newcommand{\oi}{O\thinspace{\sc i}}
\newcommand{\oii}{O\thinspace{\sc ii}}
\newcommand{\sii}{S\thinspace{\sc ii}}
\newcommand{\oiii}{O\thinspace{\sc iii}}
\newcommand{\hii}{H\thinspace{\sc ii}}

\begin{abstract}

We analyze deep JWST/NIRSpec observations of 58 star-forming galaxies at $1.3< z <4.8$ in the AURORA survey to investigate the origin of the permitted O\thinspace{\sc i} $\lambda 8449$ emission feature. Based on the detection of O\thinspace{\sc i} $\lambda 8449$ in a stack of objects with non-detections, we infer that this feature is ubiquitous among star-forming galaxies at $\sim$0.5\% the strength of H$\alpha$. We additionally compare our results with measurements of this line in local H\thinspace{\sc ii} regions from the CHAOS survey and star-forming galaxies at $z<0.1619$ in the DESI survey, finding a similar strength of \oi\ $\lambda 8449$ relative to H$\alpha$ in these objects. We evaluate four production mechanisms: recombination, collisional excitation, Ly$\beta$ fluorescence, and stellar continuum fluorescence. Of these mechanisms, stellar continuum fluorescence (rather than Ly$\beta$ fluorescence) provides the most compelling explanation for the O\thinspace{\sc i} $\lambda 8449$ emission based on the additional detection of the near-infrared O\thinspace{\sc i} $\lambda 11290$ and O\thinspace{\sc i} $\lambda 13168$ emission lines. Adopting physical conditions derived from a stacked composite spectrum, we make predictions for the contribution of the other mechanisms to O\thinspace{\sc i} $\lambda 8449$, determining that recombination, collisional excitation, and Ly$\beta$ fluorescence contribute negligibly to this feature. While a simple H\thinspace{\sc ii} region model using {\sc Cloudy} reproduces the average observed O\thinspace{\sc i} $\lambda 8449$/H$\alpha$ ratios in the AURORA sample, future works focused on refining model density profiles will be a valuable step in better explaining the near-infrared \oi\ emission-line strengths.
\end{abstract}



\section{Introduction}

The near-infrared (NIR) sensitivity of JWST has vastly improved our ability to probe the conditions in the interstellar medium (ISM) of star-forming galaxies using rest-optical features at $z\gtrsim2$. While a great deal of work has examined the properties of galaxies using the brightest rest-optical emission lines (e.g., H$\alpha$, [\oiii] $\lambda\lambda 4960,5008$), a wealth of information is encoded in the intrinsically faint emission lines that are now accessible with deep JWST surveys. 

One such faint emission feature is the \oi\ $\lambda 8449$ triplet $\rm (3s\ ^3S_1- 3p\ ^3P)$, which has now been detected in the spectra of active galactic nuclei (AGN) and Little Red Dots (LRDs) with JWST \citep[e.g.,][]{2024MNRAS.535..853J,2024arXiv241204557L,2025arXiv250921575U,2025ApJ...994L...6T,2025arXiv251121820D,2025ApJ...984..121W,2026ApJ..1004..153K,2026MNRAS.545f2117D,2026arXiv260403563T,2026arXiv260403370J} as well as in some star-forming galaxies \citep{2023ApJ...958L..11S,2026A&A...710A..18C}. \oi\ $\lambda 8449$ is a relatively common feature of AGN spectra \citep[e.g.,][]{1980ApJ...238...10G,2008ApJS..174..282L,2024A&A...686A..17O}, and is present in a variety of other astrophysical sources such as supernovae/supernova remnants \citep[e.g.,][]{1985ApJ...299..981W,2019A&A...628A..93P,2017MNRAS.469.1559G}, planetary nebulae \citep[e.g.,][]{1989ApJ...346..799R}, X-ray binaries, Galactic novae \citep[e.g.,][]{2024MNRAS.532.3985P}, emission-line stars \citep[e.g.,][]{1947ApJ...105..212H,2012ApJ...753...13M,2018ApJ...857...30M}, the Orion nebula \citep[e.g.,][]{1971MNRAS.153..393M,1975ApJ...196..465G}, a nearby starburst galaxy \citep{1995ApJ...439..604G}, and ``Godzilla" in the Sunburst Arc \citep{2025A&A...698A..16C}. 

\begin{figure*}
    \centering
    \includegraphics[width=\textwidth]{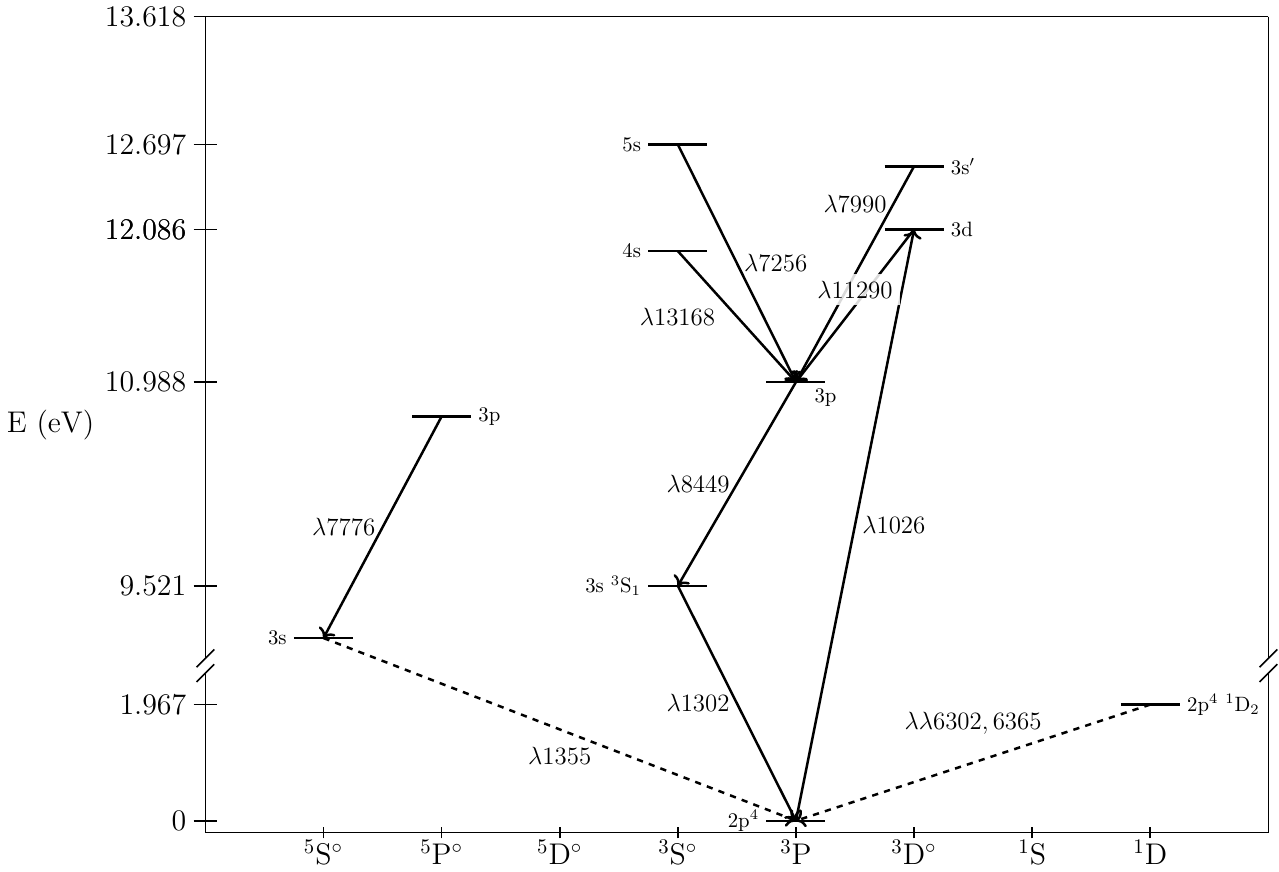}
    \caption{A simplified \oi\ energy level diagram showing the most relevant energy levels and transitions discussed in this work.}
    \label{fig:energy_level_diagram}
\end{figure*}

The production of \oi\ $\lambda 8449$ is commonly interpreted in the context of Ly$\beta$ or ``Bowen" fluorescence, due to a photoexcitation by accidental resonance (PAR) between the hydrogen Ly$\beta$ line and the \oi\ $\lambda 1026$ transition. In this process, a Ly$\beta$ photon excites the 3d $\rm ^3D^\circ$ level of \oi, leading to the subsequent emission of an \oi\ $\lambda 11290$, $\lambda 8449$, and $\lambda 1302$ photon. A simplified energy-level diagram for \oi\ is shown in Figure \ref{fig:energy_level_diagram} for reference. The \oi\ $\lambda 1026$ line is illustrated as an absorption transition to highlight the Ly$\beta$ PAR pathway. 

The Ly$\beta$ fluorescence mechanism is the favored explanation for \oi\ emission in the broad-line regions of AGN, where the gas column density is extremely high \citep{1947PASP...59..196B,1980ApJ...238...10G,1995ApJS...96..325B,2002ApJ...572...94R}. If one assumes PAR with Ly$\beta$ is responsible for \oi\ $\lambda 8449$, its discovery in star-forming galaxies is puzzling, given that the required conditions for this process (a dense medium optically thick to H$\alpha$, \citealt{1980ApJ...238...10G}) are not typically satisfied in star-forming regions. Another commonly discussed process is collisional excitation. This process has been shown to contribute appreciably to the production of \oi\ $\lambda 8449$ in the spectra of Seyfert 1 galaxies \citep{2002ApJ...572...94R}, but its role in the production of this line in star-forming regions is uncertain. \oi\ $\lambda 8449$ is also discussed as a feature of star-forming regions, arising from the recombination of free electrons with O$^+$ \citep[e.g.,][]{2019A&ARv..27....3M}. The detection of \oi\ $\lambda8449$ as a recombination line in star-forming galaxies is also unexpected, given its strength relative to H$\alpha$ in these objects should be too small to yield a detection even with some of the deepest existing integrations \citep[e.g.,][]{1975ApJ...196..465G}.

Another compelling explanation for the production of \oi\ $\lambda 8449$ is stellar continuum fluorescence. This mechanism has been presented as the preferred explanation for the strength of \oi\ $\lambda 8449$ in the Orion nebula \citep{1975ApJ...196..465G} as well as several planetary nebulae \citep[e.g.,][]{1992ApJ...384..536R}. However, distinguishing between PAR with Ly$\beta$ and stellar continuum fluorescence has proven challenging in observations of high-redshift galaxies due to a lack of measurements of two key NIR emission lines: \oi\ $\lambda 11290$ ($\rm 3p\  ^3 P- 3d\ ^3D^\circ$) and \oi\ $\lambda 13168$ ($\rm 3p\  ^3 P- 4s\ ^3S^\circ$). The lower levels of each of these transitions correspond to the upper level of the \oi\ $\lambda 8449$ transition, making them valuable probes of the mechanism that excites \oi\ $\lambda 8449$. However, the \oi\ $\lambda 11290$ line is emitted as part of the Ly$\beta$ PAR process, while \oi\ $\lambda 13168$ is not, meaning that the presence/absence of each of these lines places strong constraints on possible \oi\ fluorescence mechanisms.

In this study, we report the detection of \oi\ $\lambda 8449$ in 14 star-forming galaxies from the AURORA survey in the range $1.8<z<4.4$, with a median redshift of $z_{\rm med} = 2.32$. For the first time at high redshift, we utilize measurements of the \oi\ $\lambda 11290$ and \oi\ $\lambda 13168$ lines to constrain the production mechanism of \oi\ $\lambda 8449$ in the star-forming galaxy population.

In Section \ref{sec:observations}, we describe the observations as well as the construction of composite ``stacked" spectra. In Section \ref{sec:results}, we describe the results of our stacking analysis and provide an overview of the sample properties. In Section \ref{sec:excitation}, we explore the contributions of various excitation mechanisms to the \oi\ $\lambda 8449$ emission. In Section \ref{sec:discussion}, we discuss how our findings affect the interpretation of this line in star-forming galaxy spectra. Throughout this analysis, we adopt a \citet{2003PASP..115..763C} initial mass function (IMF), a Hubble constant of $H_0=\rm 70\ km\ s^{-1}\ Mpc^{-1}$, $\Omega_\Lambda =0.7$, $\Omega_m=0.3$, and solar chemical abundances of $\rm 12+\log(O/H)_\odot = 8.69$ and $Z_\odot=0.014$ \citep{2021A&A...653A.141A}.


\section{Observations} \label{sec:observations}

\subsection{The AURORA survey}\label{sec:survey}

The spectroscopic observations on which this analysis is based come from the Cycle 1 GO program AURORA \citep{2025ApJ...980..242S}, a JWST/NIRSpec micro-shutter array (MSA) survey of 97 objects with $z_{\rm med} = 2.57$, characterized by very deep integrations of 12.3 h, 8.0 h, and 4.2 h in the G140M, G235M, and G395M $R\sim 1000$ gratings, respectively. Further details regarding the data reduction, slit loss corrections, and flux calibrations can be found in \citet{2025ApJ...980..242S}. Briefly, the two-dimensional (2D) NIRSpec spectra were reduced using a slightly modified version of the standard STScI data reduction pipeline routines \citep[v1.13.4;][]{bushouse_2024_10569856}. One-dimensional (1D) spectra were then extracted from the 2D frames and corrected for slit losses as described in \citet{2026ApJ...999...15R}. The spectra in each grating were then scaled relative to each other according to the procedure described by \citet{2025ApJ...989..209S} before being multiplicatively scaled to match the available JWST/NIRCam photometry.

\begin{figure*}
    \centering
    \includegraphics[width=\linewidth]{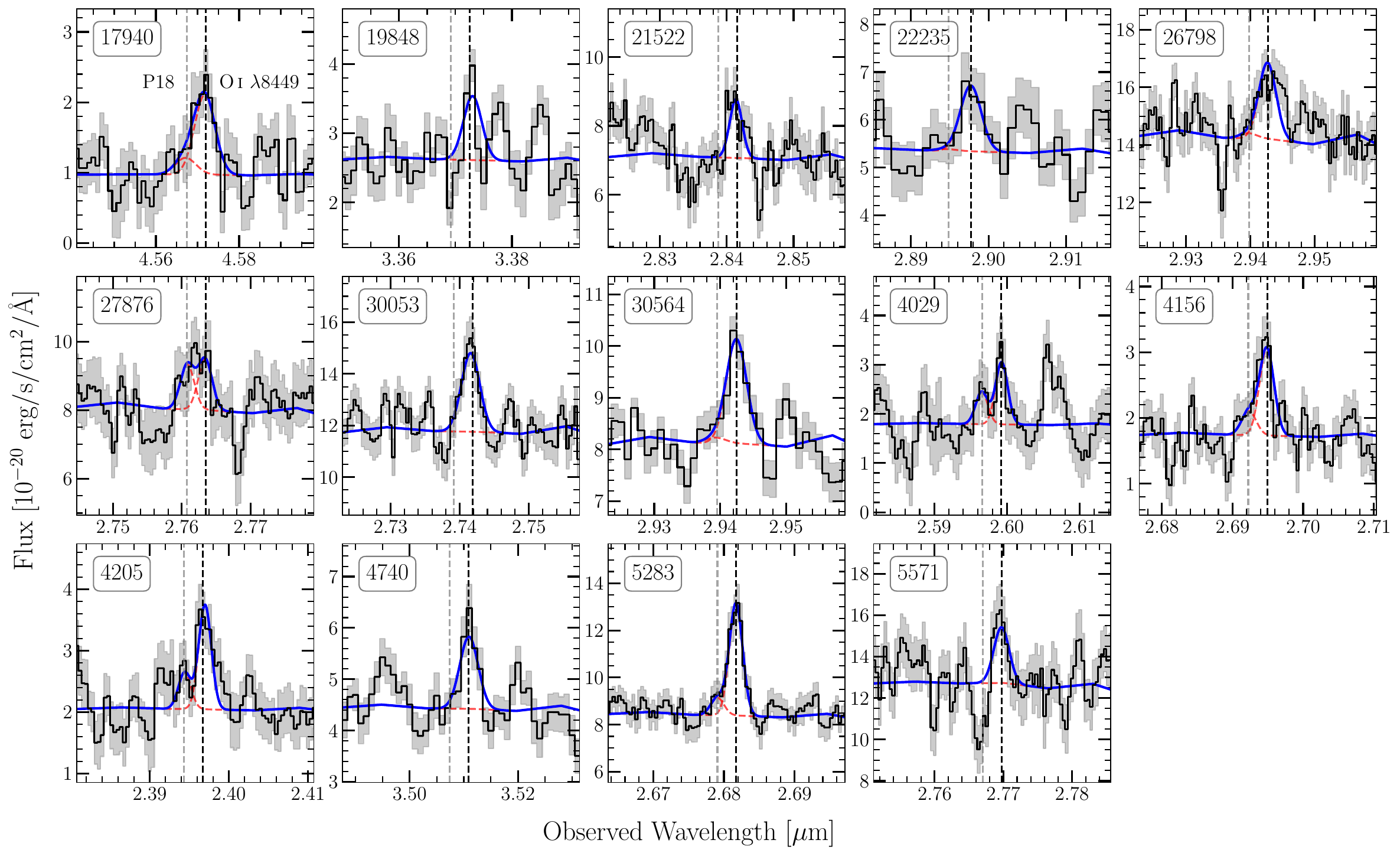}
    \caption{Individual \oi\ $\lambda 8449$ detections in the AURORA sample. The vertical black dashed line shows the position of \oi\ $\lambda 8449$, and the gray dashed line shows the position of Paschen 18. The full line profile is shown in blue, while the individual line components are shown as dashed red curves.}
    \label{fig:individual_fits}
\end{figure*}

\subsection{Emission-line measurements and the \oi\ sample}\label{sec:emline_meas}

Emission lines were measured from the extracted, slit-loss-corrected, flux-calibrated spectra described in Section \ref{sec:survey}. Each emission line was modeled with a Gaussian profile. In order to correct for underlying stellar absorption features (most important for measuring the Balmer lines), all lines were fit with a Gaussian profile, using the best-fit spectral energy distribution (SED) as the continuum model. Object SEDs were fit using the FAST code \citep{2009ApJ...700..221K}, utilizing public JWST/NIRCam, HST/WFC3, and HST/ACS photometry corrected for nebular emission using the flux-calibrated AURORA spectra in an iterative procedure. Further details on the SED fitting procedure can be found in \citet{2025ApJ...980..242S}. As discussed by \citet{2025MNRAS.541.1707T}, fitting the SEDs using {\sc Prospector} \citep{2021ApJS..254...22J} produced similar results to those obtained with FAST. The line fitting was performed using the \texttt{scipy.optimize.curve\_fit()} function, which utilizes the Trust Region Reflective algorithm for optimization with bounds \citep{branch1999subspace}. Uncertainties on the line fluxes were obtained through Monte Carlo simulations, perturbing the science spectra by their respective uncertainties and re-fitting 1000 times, and taking the standard deviation of these simulations to be the 1$\sigma$ uncertainties on the line fluxes. For emission lines appearing in the overlap region of two gratings, the line was fit in each grating, and the inverse-variance-weighted average flux of the line was adopted as the fiducial value. Among the AURORA galaxies, we identified 16 objects as having $>$5$\sigma$ detections of the \oi\ $\lambda 8449$ line. One object (ID 921842) was identified as an AGN due to the presence of broad Balmer emission, and was thus removed from the analysis. Another object (ID 6124) was reported to have a $>$5$\sigma$ detection of \oi\ $\lambda 8449$, but the detection was ruled as unconvincing based on visual inspection of the science spectrum, and was thus removed from the sample of individual detections. Thus, we present a sample of 14 detections of \oi\ $\lambda 8449$ in individual star-forming galaxies. The emission-line fits of \oi\ $\lambda 8449$ (partly blended with Paschen 18) can be seen in Figure \ref{fig:individual_fits}.

\subsection{Composite spectra}
In addition to individual objects, we also analyze the composite stacked spectra of objects in the AURORA sample. We created stacks in two configurations, one stacking together objects with $>$5$\sigma$ detections of \oi\ $\lambda 8449$, and another with coverage of the \oi\ $\lambda 8449$ line, but no detection. We refer to these stacks as {\sc Stack-Det} and {\sc Stack-NonDet}, respectively (Object ID 6124, which was removed from the {\sc Stack-Det} sample as mentioned in Section \ref{sec:emline_meas}, was included in the {\sc Stack-NonDet} sample). The {\sc Stack-Det} and {\sc Stack-NonDet} composites are composed of 14 and 44 star-forming galaxies with median redshifts of $z_{\rm med}=2.32$ and $z_{\rm med}=2.31$, median stellar masses of $\log(M_*/M_\odot)=9.48\pm0.06$ and $\log(M_*/M_\odot)=9.18\pm0.06$, and median star-formation rates of $\log({\rm SFR}/ M_\odot\ {\rm y^{-1}}) = 1.68\pm0.01$ and $\log({\rm SFR}/ M_\odot\ {\rm y^{-1}}) = 0.69\pm0.02$, respectively. These median stack properties are tabulated in Table \ref{tab:line_ratios}.

When combining spectra, we first subtracted off the stellar continuum emission using the best-fit SED as a model, multiplying this SED by a cubic polynomial to fit the shape of the grating spectrum over pre-defined continuum windows prior to subtraction. After subtraction, we corrected each spectrum for dust attenuation using an individually determined dust curve as described in \citet{2026ApJ...999...15R}. For cases where an individual dust curve was not able to be determined, the average nebular dust curve from \citet{2026ApJ...999...15R} was used. For this dust correction, we assumed an intrinsic Balmer decrement of H$\alpha$/H$\beta=2.79$, corresponding to $T_e=15,000$ K and $n_e=\rm 100\ cm^{-3}$ for Case B recombination.

Each continuum-subtracted, dust-corrected spectrum was then normalized to the H$\alpha$ luminosity, shifted into the rest frame, and interpolated onto a common wavelength grid, where the wavelength spacing at each pixel $\Delta \lambda$ was calculated to sample the effective stacked line-spread function (LSF) with one pixel per LSF width ($\sigma_{LSF}$). The effective $\sigma_{LSF}$ of each stack was determined by calculating the median value of the spectral resolution $R$ at each rest-frame wavelength across all constituent galaxies of that stack. The resolution curves for each galaxy were modeled based on the available JWST/NIRCam imaging using the \texttt{msafit} package \citep{2024A&A...684A..87D}. A detailed description of this procedure can be found in \citet{2025ApJ...989..209S}. Once the median, effective resolution curve $R_{med}$ was determined for a given stack, the spacing of the wavelength grid, $\Delta \lambda$, was set equal to $\sigma_{LSF}$, where $\Delta\lambda = \sigma_{LSF}=\lambda/(2.355R_{med})$. Using the {\sc SpectRes} package \citep{2017arXiv170505165C}, we then resampled each individual spectrum onto the common grid defined by the wavelength-dependent $\Delta \lambda$ values.

\begin{figure*}
    \centering
    \includegraphics[width=\linewidth]{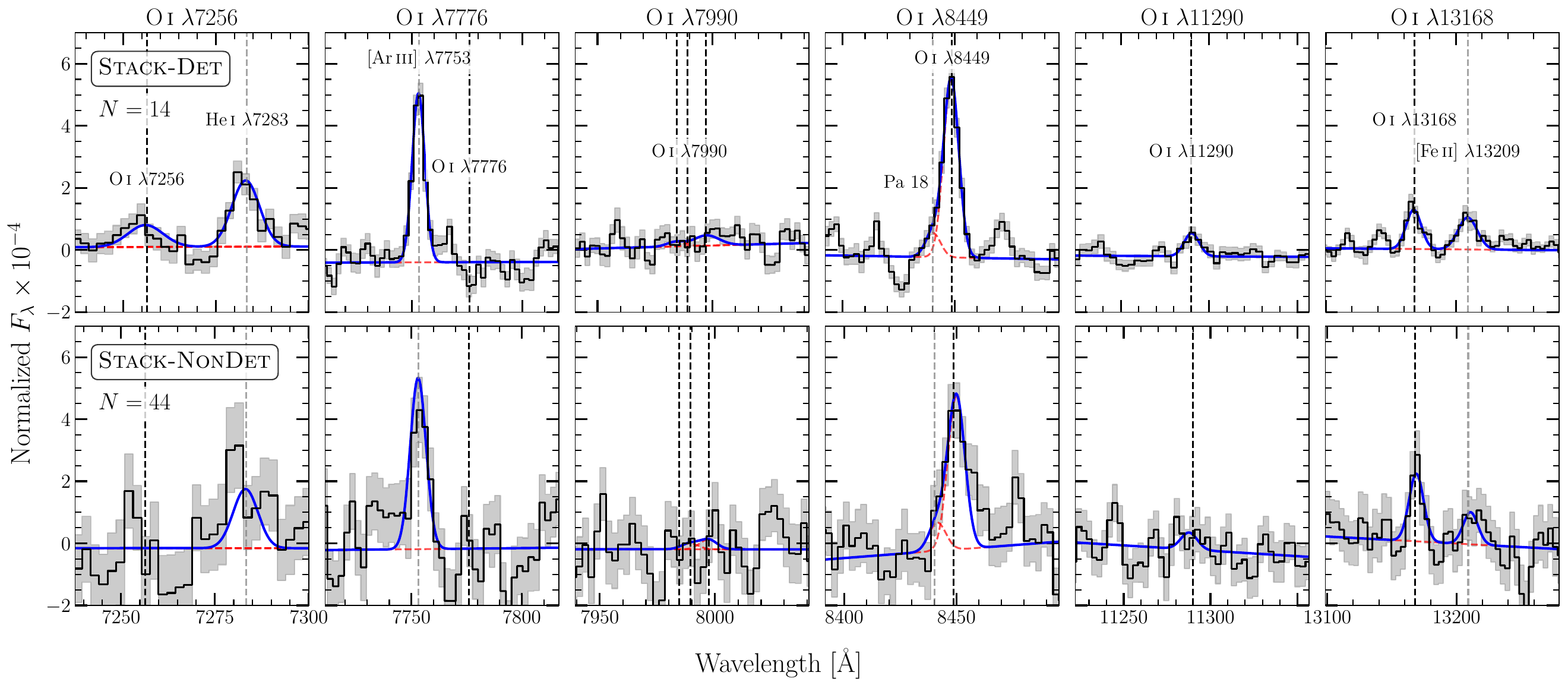}
    \caption{Stack \oi\ line fits. The vertical black dashed lines show the positions of each \oi\ line. The full line profile is shown in blue, while the individual line components and continuum models are shown as dashed red curves.}
    \label{fig:stacked_fits}
\end{figure*}

The set of resampled spectra was subsequently median combined to create a composite spectrum. To determine the uncertainty at each wavelength, we performed 1000 Monte Carlo simulations, each time perturbing each component science spectrum by its respective error at each pixel, perturbing the value of $E(B-V)$ for each spectrum by its uncertainty, and performing a bootstrap resampling with replacement to capture sample variance. The median and standard deviation of these simulations represent the flux density and the error at each pixel, respectively. Finally, we rescaled the stacked error spectrum by measuring the RMS fluctuations of the flux density in pre-defined continuum windows, multiplying the error spectrum by a constant factor such that the median value of the error spectrum was equal to this RMS value. We then fit the emission lines in the stacked spectra in a manner similar to that described in Section \ref{sec:emline_meas}, this time using a linear continuum model since the continuum contributions (e.g., stellar Balmer absorption) were removed from the science spectra prior to stacking. Fits to several key \oi\ emission lines in the {\sc Stack-Det} and {\sc Stack-NonDet} composite spectra are shown in Figure \ref{fig:stacked_fits}. The IDs, redshifts, and normalized emission-line strengths for the individual objects as well as the stacks are compiled in Table \ref{tab:line_ratios}. In cases of non-detections, 3$\sigma$ upper limits are indicated.

\section{Results}\label{sec:results}

\begin{figure}
    \centering
    \includegraphics[width=8.8cm]{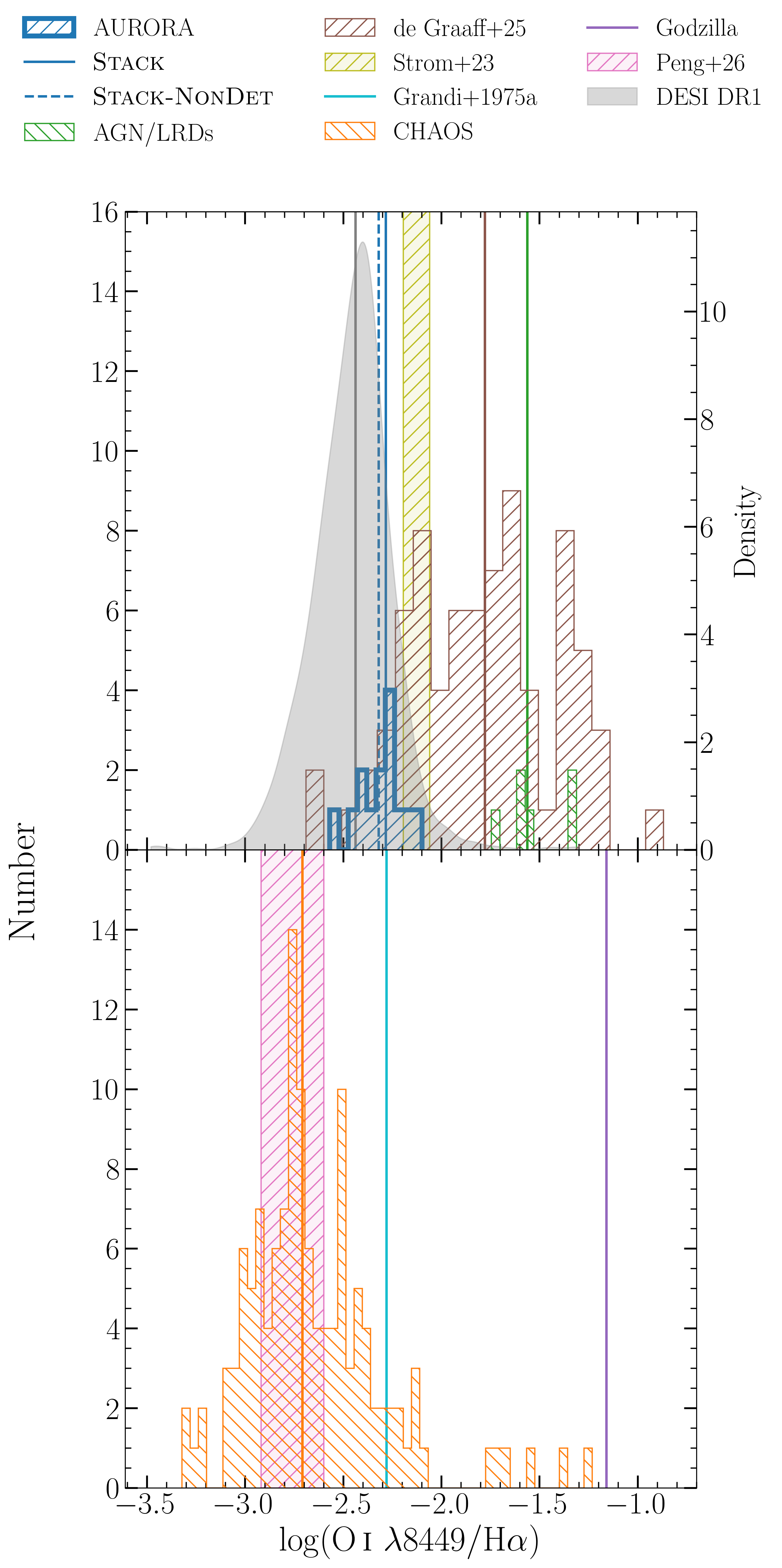}
    \caption{Comparison of the \oi\ $\lambda 8449$/H$\alpha$ ratio among several extragalactic samples. ({\it Top}): Galaxies from AURORA and DESI, and a compilation of AGN and LRD observations from JWST \citep{2024MNRAS.535..853J, 2024arXiv241204557L, 2025arXiv251121820D,2025arXiv250921575U, 2026ApJ..1004..153K, 2026arXiv260403563T}. ({\it Bottom}): Resolved \hii\ regions from the CHAOS survey and SSCs such as ``Godzilla" in the Sunburst Arc \citep{2025A&A...698A..16C} and in SBS 0335-052 E \citep{2026ApJ..1001..218P}.}
    \label{fig:oi_ha}
\end{figure}

We present detections of \oi\ $\lambda 8449$ in 14 galaxies from the AURORA survey, as well as detections in two spectral stacks comprising a total of 58 star-forming galaxies. As seen in Figure \ref{fig:stacked_fits}, there is a clear detection of \oi\ $\lambda 8449$ in both the {\sc Stack-Det} composite spectrum and the {\sc Stack-NonDet} composite, indicating that the presence of \oi\ $\lambda 8449$ emission (at the level of $\sim 5\times 10^{-3}$ times the strength of H$\alpha$) is ubiquitous among star-forming galaxies. This emission has only been recently observed in $z\sim 2$ star-forming galaxies as a result of surveys that are characterized by very deep integration times \citep[e.g., CECILIA, MARTA;][]{2023ApJ...958L..11S,2026A&A...710A..18C}. Of the other permitted \oi\ lines that we measure in this analysis, the strongest line after \oi\ $\lambda 8449$ is, in most cases, the \oi\ $\lambda 13168$ line, yielding a $>$3$\sigma$ detection in both stacks as well as 6/10 individual objects with coverage of this feature. The \oi\ $\lambda 11290$ line, by contrast, only yields a detection in 3 out of 13 objects with coverage, is not detected in the {\sc Stack-NonDet} composite, and is measured to be $42\pm10\%$ the strength of \oi\ $\lambda13168$ in the {\sc Stack-Det} composite. In the Stack-NonDet composite, the \oi\ $\lambda 11290$ line is not detected (formally at $<$50\% the strength of \oi\ $\lambda 13168$). Two other \oi\ lines, $\lambda7990$ and $\lambda 7776$, often used as probes of recombination and stellar continuum fluorescence \citep[e.g.,][]{1980ApJ...238...10G}, are not detected in any individual object or either of the stacks. \oi\ $\lambda 7256$, another feature analyzed in nearby star-forming regions \citep[e.g.,][]{1975ApJ...196..465G,1975ApJ...199L..43G}, yields a robust detection in three objects: CO-4205, CO-5283, and GN-26798.\footnote{We note the presence of a spurious feature at $\lambda = 7267\rm \AA$ in the G395M spectrum of GN-19848 that our line fitting code incorrectly attributes to \oi\ $\lambda 7256$.}

The strength of the \oi\ $\lambda 8449$ line can be characterized by the \oi\ $\lambda 8449$/H$\alpha$ ratio, tabulated for this sample in Table \ref{tab:line_ratios}, and illustrated as a histogram in Figure \ref{fig:oi_ha}. The median log(\oi\ $\lambda 8449$/H$\alpha$) ratio from the sample of \oi\ $\lambda 8449$ detections in AURORA is $-2.29\pm0.03$, while the log(\oi\ $\lambda 8449$/H$\alpha$) ratio measured from the {\sc Stack-NonDet} composite is $-2.32\pm 0.05$. These star-forming galaxies are characterized by an \oi\ $\lambda 8449$ strength that spans a range of 0.27\% to 0.80\% the strength of H$\alpha$. This range is comparable to findings by \citet{2023ApJ...958L..11S} based on a stack of 23 $z\sim 2$ galaxies, shown in the top panel of Figure \ref{fig:oi_ha}.

The measurement of log(\oi\ $\lambda 8449$/H$\alpha)=-2.28$ in the Orion nebula made by \citet{1975ApJ...196..465G} is shown in the bottom panel of Figure \ref{fig:oi_ha}. We note that we have converted his original measurement of \oi\ $\lambda 8449$/H$\beta$ to \oi\ $\lambda 8449$/H$\alpha$, adopting H$\alpha$/H$\beta=2.86$ as a conversion, which corresponds to $T_e=10,000$ K and $n_e=100\rm\ cm^{-3}$. This \oi\ $\lambda 8449$/H$\alpha$ ratio is nearly identical to the median value from the AURORA sample. This similarity is an indicator that the physical process that produces \oi\ $\lambda 8449$ emission in Orion may also be at work in the AURORA galaxies.

We also compare the AURORA \oi\ $\lambda 8449$/H$\alpha$ ratios with those measured from local \hii\ regions drawn from the CHAOS survey \citep{2015ApJ...806...16B,2015ApJ...808...42C,2020ApJ...893...96B,2021ApJ...915...21R,2022ApJ...939...44R,2024ApJ...971...87B}. The median value of log(\oi\ $\lambda 8449$/H$\alpha)$ for the sample of CHAOS \hii\ regions is $-2.71\pm0.01$. This value, illustrated in the bottom panel of Figure \ref{fig:oi_ha} as a vertical orange line, is lower than the median log(\oi\ $\lambda 8449$/H$\alpha$) of the AURORA detections by $0.42\pm0.03$ dex. Even when comparing the CHAOS median log(\oi\ $\lambda 8449$/H$\alpha$) with the log(\oi\ $\lambda 8449$/H$\alpha$) ratio measured from {\sc Stack-NonDet}, the two ratios differ by a similar value, $0.39\pm0.05$ dex. Because the \oi\ $\lambda 8449$/H$\alpha$ ratios in the objects with detections are very similar to the \oi\ $\lambda 8449$/H$\alpha$ ratio measured in a stack of non-detections, this similarity suggests that the sample with \oi\ $\lambda 8449$ detections is not biased toward high \oi\ $\lambda 8449$/H$\alpha$ ratios. Furthermore, this similarity suggests that the observed difference between CHAOS and AURORA is not driven by a bias in the AURORA sample toward bright \oi\ $\lambda 8449$ emitters. Because the CHAOS apertures are centered on local \hii\ regions and are dominated by ionized gas emission, it is possible that the lower observed \oi\ $\lambda 8449$/H$\alpha$ in CHAOS compared to AURORA is an aperture effect. \oi\ emission in the Orion Bar, for example, has been observed to arise from the ionization front and the surrounding neutral photodissociation region \citep{2024A&A...685A..74P}. If this \oi\ spatial distribution characterizes all \hii\ regions, then small, \hii-region-centered slit apertures such as those used in CHAOS may miss a portion of the \oi\ emission and give rise to the \oi\ $\lambda 8449$/H$\alpha$ discrepancy that we measure. 

The \oi\ $\lambda 8449$ line was also recently observed in the blue compact dwarf galaxy SBS 0335-052 E by \citet{2026ApJ..1001..218P}. Though the authors interpret this emission in the context of Ly$\beta$ fluorescence, we note that their range of observed \oi\ $\lambda 8449$/H$\alpha$ values for the 8 super star clusters (SSCs) in this object is in close agreement with the range of \oi\ $\lambda 8449$/H$\alpha$ values from the CHAOS survey.

We additionally compare our measured \oi\ $\lambda 8449$/H$\alpha$ ratios with a sample of 1602 galaxies from data release 1 of the Dark Energy Spectroscopic Instrument (DESI) survey \citep{2026AJ....171..285D}. To assemble this comparison sample, we restricted the full DESI parent sample to galaxies at $z<0.1619$ to ensure \oi\ $\lambda 8449$ coverage, required a signal-to-noise ratio (SNR) $>$100 on H$\alpha$, and SNR$>$5 on \oi\ $\lambda 8449$. AGN were removed using the \citet{2003MNRAS.346.1055K} delineation, and all line fluxes were corrected for dust using the Balmer decrement, assuming a \citet{1989ApJ...345..245C} dust law and H$\alpha$/H$\beta=2.86$, corresponding to $T_e=10,000$ K and $n_e = 100\rm \ cm^{-3}$. The kernel density estimation (KDE) of the DESI sample is shown in gray in Figure \ref{fig:oi_ha}. The median value of log(\oi\ $\lambda 8449$/H$\alpha)=-2.447\pm0.003$ for the DESI sample lies $0.16\pm0.03$ dex lower than the median of the AURORA \oi\ $\lambda 8449$/H$\alpha$ distribution. This value places the DESI galaxies closer to the distribution of AURORA galaxies than the CHAOS \hii\ regions. Because the DESI measurements are made on galaxy-integrated spectra, the fact that the DESI \oi\ $\lambda 8449$/H$\alpha$ distribution more closely matches AURORA than CHAOS is consistent with the CHAOS \oi\ $\lambda 8449$/H$\alpha$ spectra sampling a relatively small fraction of the neutral ISM.

In addition to star-forming sources such as the CHAOS \hii\ regions and low-$z$ DESI galaxies, we also compare with samples of LRDs and AGN that have been observed with JWST \citep{2024MNRAS.535..853J, 2024arXiv241204557L, 2025arXiv251121820D, 2025arXiv250921575U, 2026ApJ..1004..153K, 2026arXiv260403563T}. Because these ratios are not corrected for dust attenuation, the AGN \oi\ $\lambda 8449$/H$\alpha$ ratios represent upper limits. For the LRDs, significant dust attenuation may not contribute, given that dust emission is typically not detected in this class of objects \citep[e.g.,][]{2024ApJ...975L...4C,2025ApJ...992...26L,2025ApJ...991...37A,2025ApJ...978...92L}. With this caveat in mind, we note that the median value of the distribution of log(\oi\ $\lambda 8449$/H$\alpha$) in AGN/LRDs lies $0.73\pm 0.03$ dex higher than the median of the AURORA \oi\ $\lambda 8449$/H$\alpha$ detections. Additionally, the median \oi\ $\lambda 8449$/H$\alpha$ value from the compilation of LRDs from \citet{2025arXiv251121820D} lies $0.51\pm 0.03$ dex higher than the AURORA sample median. These differences may arise as a result of a different excitation mechanism for \oi\ $\lambda 8449$ compared to the star-forming galaxies.

Finally, we note the extreme nature of the ``Godzilla" object in the Sunburst Arc presented by \citet{2025A&A...698A..16C}. The \oi\ $\lambda 8449$/H$\alpha$ ratio in this object stands out compared to all classes of object that we analyze in this study, likely indicating unique and extreme conditions giving rise to the \oi\ emission in this object. Indeed, \citet{2025A&A...698A..16C} draw parallels between the conditions in Godzilla and the Weigelt blobs, a group of dense gas clouds surrounding the central star of $\eta$ Car \citep{1986A&A...163L...5W,2005MNRAS.364..731J}. As such, the strong \oi\ emission in this object substantially sets it apart from our $z\sim 2$ star-forming galaxy sample. Overall, the wide range of \oi\ $\lambda 8449$/H$\alpha$ ratios in various extragalactic astrophysical sources suggests a variety of excitation mechanisms of \oi\ $\lambda 8449$ arising from vastly different conditions. Notably, the \oi\ $\lambda 8449$/H$\alpha$ ratios among the AURORA, CECILIA, and DESI samples, along with Orion, are very similar, suggesting that they share a common \oi\ excitation mechanism. This argument may also apply to the CHAOS \hii\ regions if aperture effects contribute significantly to their lower \oi\ $\lambda 8449$/H$\alpha$ ratios. In the following Section, we explore the possible excitation mechanisms for \oi\ $\lambda 8449$ in our sample of star-forming galaxies.

\section{Determining the \oi\ excitation mechanism} \label{sec:excitation}

\begin{deluxetable*}{cccccccccccc}
\tabletypesize{\scriptsize}
\label{tab:line_ratios}
\caption{Emission-line strengths normalized to \oi\ $\lambda 8449$.}
\tablehead{
 \colhead{ID} & \colhead{$z$} & \colhead{$\log(M_*/M_\odot)$} & \colhead{log(SFR/$M_\odot\ \rm yr^{-1}$)} & \colhead{$E(B-V)$} & \colhead{\oi\ $\lambda 8449$/H$\alpha^*$} & \colhead{[\oi]\ $\lambda 6302$} & \colhead{$\lambda 7256$} & \colhead{$\lambda 7776$} & \colhead{$\lambda 7990$} & \colhead{$\lambda 11290$} & \colhead{$\lambda 13168$}}\startdata
\hline
CO-4205 & $1.8370$ & $8.33^{+0.01}_{-0.00}$ & $1.11^{+0.01}_{-0.01}$ & $0.11^{+0.00}_{-0.00}$ & $(4.0\pm0.5)\times $$10^{-3}$ & $5.55\pm0.74$ & $0.42\pm0.12$ & $<0.48$ & $<0.73$ & $<0.56$ & $0.44\pm0.12$ \\
CO-4029 & $2.0765$ & $8.44^{+0.00}_{-0.07}$ & $1.30^{+0.01}_{-0.01}$ & $0.09^{+0.01}_{-0.01}$ & $(2.7\pm0.7)\times $$10^{-3}$ & $5.76\pm1.40$ & $<0.73$ & $<0.97$ & -- & $<0.90$ & $0.91\pm0.31$ \\
CO-5283 & $2.1740$ & $9.47^{+0.03}_{-0.11}$ & $1.63^{+0.01}_{-0.01}$ & $0.08^{+0.00}_{-0.00}$ & $(5.7\pm0.5)\times $$10^{-3}$ & $3.65\pm0.32$ & $0.35\pm0.07$ & $<0.17$ & $<0.38$ & $<0.25$ & $<0.22$ \\
CO-4156 & $2.1897$ & $8.66^{+0.09}_{-0.03}$ & $1.30^{+0.01}_{-0.01}$ & $0.08^{+0.00}_{-0.00}$ & $(3.6\pm0.5)\times $$10^{-3}$ & $3.24\pm0.45$ & $<0.36$ & $<0.31$ & $<0.58$ & $<0.52$ & $<0.45$ \\
GN-30053 & $2.2450$ & $9.90^{+0.05}_{-0.26}$ & $1.96^{+0.01}_{-0.01}$ & $0.41^{+0.01}_{-0.01}$ & $(5.6\pm0.9)\times $$10^{-3}$ & $7.22\pm0.85$ & $<0.39$ & $<0.29$ & $<0.78$ & $<0.20$ & $0.22\pm0.06$ \\
GN-27876 & $2.2709$ & $10.19^{+0.02}_{-0.40}$ & $1.73^{+0.02}_{-0.02}$ & $0.47^{+0.01}_{-0.01}$ & $(4.3\pm1.8)\times $$10^{-3}$ & $3.95\pm1.24$ & $<0.94$ & $<0.72$ & $<1.51$ & $<0.60$ & $0.47\pm0.19$ \\
CO-5571 & $2.2783$ & $10.26^{+0.02}_{-0.01}$ & $1.49^{+0.02}_{-0.02}$ & $0.22^{+0.01}_{-0.01}$ & $(5.0\pm1.4)\times $$10^{-3}$ & $6.27\pm1.51$ & $<0.55$ & $<0.61$ & $<1.13$ & $<0.82$ & $<0.81$ \\
GN-21522 & $2.3632$ & $9.49^{+0.11}_{-0.06}$ & $1.07^{+0.02}_{-0.02}$ & $0.01^{+0.01}_{-0.01}$ & $(6.5\pm1.2)\times $$10^{-3}$ & $3.51\pm0.73$ & $<0.59$ & $<0.55$ & $<0.98$ & $<0.56$ & $0.98\pm0.28$ \\
GN-22235 & $2.4296$ & $9.16^{+0.12}_{-0.05}$ & $1.50^{+0.02}_{-0.02}$ & $0.14^{+0.01}_{-0.01}$ & $(5.5\pm0.7)\times $$10^{-3}$ & $4.03\pm0.52$ & $<0.27$ & $<0.37$ & -- & $0.48\pm0.10$ & -- \\
GN-30564 & $2.4826$ & $9.93^{+0.12}_{-0.11}$ & $1.95^{+0.01}_{-0.01}$ & $0.40^{+0.01}_{-0.01}$ & $(6.4\pm0.7)\times $$10^{-3}$ & $4.73\pm0.40$ & $<0.30$ & $<0.22$ & $<0.51$ & $<0.17$ & $<0.16$ \\
GN-26798 & $2.4829$ & $10.49^{+0.00}_{-0.04}$ & $1.92^{+0.02}_{-0.02}$ & $0.26^{+0.01}_{-0.01}$ & $(8.0\pm1.5)\times $$10^{-3}$ & $8.24\pm1.18$ & $0.42\pm0.15$ & $<0.45$ & $<0.96$ & $0.41\pm0.12$ & $0.40\pm0.10$ \\
GN-19848 & $2.9922$ & $9.19^{+0.05}_{-0.04}$ & $1.87^{+0.04}_{-0.04}$ & $0.12^{+0.01}_{-0.01}$ & $(4.9\pm1.1)\times $$10^{-3}$ & $3.11\pm0.68$ & $0.89\pm0.29^\dagger$ & $<0.56$ & $<1.95$ & $<0.42$ & -- \\
CO-4740 & $3.1557$ & $10.05^{+0.09}_{-0.47}$ & $2.31^{+0.02}_{-0.02}$ & $0.28^{+0.01}_{-0.01}$ & $(5.4\pm1.5)\times $$10^{-3}$ & $6.50\pm1.22$ & $<0.75$ & $<0.53$ & $<1.22$ & $0.30\pm0.11$ & -- \\
GN-17940 & $4.4113$ & $9.01^{+0.03}_{-0.03}$ & $2.52^{+0.01}_{-0.01}$ & $0.31^{+0.01}_{-0.01}$ & $(3.9\pm1.0)\times $$10^{-3}$ & $6.52\pm1.16$ & $<0.41$ & $<0.42$ & $<1.27$ & -- & -- \\
{\sc Stack-Det} & 2.3209 & $9.48\pm0.06$ & $1.68\pm0.01$ & $0.18\pm0.01$ & $(5.2\pm0.2)\times $$10^{-3}$ & $5.06\pm0.26$ & $0.17\pm0.05$ & $<0.09$ & $<0.40$ & $0.16\pm0.03$ & $0.38\pm0.05$ \\
{\sc Stack-NonDet} & 2.3072 & $9.18\pm0.06$ & $0.69\pm0.02$ & $0.09\pm0.01$ & $(4.8\pm0.6)\times $$10^{-3}$ & $5.28\pm0.72$ & $<0.62$ & $<0.51$ & $<1.17$ & $<0.33$ & $0.66\pm0.15$ \\
\hline
\enddata
\tablenotetext{*}{The typical convention in the literature has been to express the strength of \oi\ $\lambda 8449$ with respect to H$\alpha$ (or H$\beta$). Thus, we express this ratio with \oi\ $\lambda 8449$ in the numerator to make for easy comparison with other works.}
\tablenotetext{\dagger}{This reported detection is likely due to a spurious feature. See Section \ref{sec:results} for details.}\end{deluxetable*}

As laid out by \citet{1980ApJ...238...10G}, four mechanisms are commonly explored to explain the origin of \oi\ $\lambda 8449$ emission: O$^+$ recombination, collisional excitation by free electrons, continuum fluorescence, and \citet{1947PASP...59..196B} fluorescence from Ly$\beta$ (also referred to as Ly$\beta$ pumping or a PAR process; \citealt{1995ApJS...96..325B}). It has been noted \citep[e.g.,][]{2002ApJ...572...94R} that several processes can simultaneously contribute to \oi\ production in an individual object, complicating the interpretation of simple diagnostics such as those presented by \citet{1980ApJ...238...10G}. With this consideration in mind, we aim to identify the dominant mechanism powering \oi\ emission in our sample by quantifying the contribution of each of these processes to the \oi\ $\lambda 8449$ feature. In this Section, we primarily compare predicted line ratios for each excitation mechanism with the emission-line measurements in the {\sc Stack-Det} composite spectrum. Firstly, its superior signal-to-noise ratio enables a robust comparison between the observed and predicted line ratios for each excitation mechanism. Secondly, the \oi\ $\lambda 8449$/H$\alpha$ ratios in both {\sc Stack-Det} and {\sc Stack-NonDet} are consistent within 1$\sigma$, strongly suggesting that the same excitation mechanism produces \oi\ $\lambda 8449$ emission in both composite samples, and thus characterizing the {\sc Stack-Det} composite is sufficient. We explore each of the four aforementioned mechanisms in the following subsections, ultimately finding that the stellar continuum fluorescence mechanism is most consistent with our observations.

\subsection{Recombination}\label{sec:recombination}
The recombination of free electrons with O$^+$ ions should produce emission-line features arising from cascades down various transition pathways of the O$^0$ atom. Such features include the \oi\ $\lambda7776$ quintet $(\rm 3s\ ^5S -3p\ ^5P)$, and not only lines produced by the transition pathways that lead to the emission of \oi\ $\lambda 8449$. In the case of recombination, the ratio of \oi\ $\lambda 7776$/\oi\ $\lambda 8449$ is expected to be roughly 5/3, based upon the ratio of the statistical weights of the \oi\ $\lambda 7776$ quintet and \oi\ $\lambda 8449$ triplet \citep{1980ApJ...238...10G}. Given that \oi\ $\lambda 7776$ emission is not detected in any of the galaxies in the AURORA sample analyzed here, and all upper limits place the \oi\ $\lambda 7776$/\oi\ $\lambda 8449$ ratio below unity, the recombination mechanism is unlikely to be a significant contributor to the \oi\ emission for any object in this sample.

We estimate the contribution of recombination to the \oi\ $\lambda 8449$ flux by using the expressions for the recombination volume emission coefficients of \oi\ $\lambda 8449$ and H$\beta$ from \citet{1975ApJ...196..465G}. From these equations, we can derive an expression for the expected ratio of \oi\ $\lambda 8449$/H$\beta$:

\begin{equation}\label{eq:recombination}
    \frac{j_{8449}}{j_{H\beta}} = \left( \frac{N_{\rm O II}}{N_{\rm HII}} \right) \times \left( \frac{\nu_{8449}}{\nu_{\rm H\beta}} \right) \times 6.84 \times T_4^{0.22}
\end{equation}

where $T_4$ is the electron temperature in units of $10^4$ K, and $N_{\rm OII}$ and $N_{\rm HII}$ are the O$^+$ and H$^+$ number densities, respectively. Using the {\sc PyNeb} package \citep{2015A&A...573A..42L}, we calculate $T_4$ and $\rm O^+/H^+$ based on the {\sc Stack-Det} composite emission-line measurements. Based on the stacked, temperature-sensitive [\oii] $\lambda\lambda 3727,3730$/[\oii] $\lambda\lambda 7322,7332$ ratio, we derive an electron temperature in the O$^+$ zone of $T_e$([\oii])$=13,696\pm750$ K, adopting a density of $n_e=442\pm88\rm\ cm^{-3}$ based on the density-sensitive [\sii] $\lambda 6733$/[\sii] $\lambda 6718$ ratio. Using these values of $T_e$ and $n_e$, we derive an O$^+$/H$^+$ abundance of $(3.603\pm 0.725)\times 10^{-5}$. Adopting these values, and assuming an intrinsic Balmer decrement of H$\alpha$/H$\beta=2.79$ to convert to a ratio with respect to H$\alpha$, we predict log(\oi\ $\lambda 8449$/H$\alpha) = -4.26\pm0.08$, more than an order of magnitude lower than the values that are typical of our sample. Taking the ratio of the predicted \oi\ $\lambda 8449$/H$\alpha$ value vs. the measured value from the stack, we estimate that recombination contributes $1.0\pm0.2$\% to the stacked \oi\ $\lambda 8449$ flux. This percentage is likely an upper limit, since the \oi\ $\lambda 8449$/H$\alpha$ ratio inferred from equation \ref{eq:recombination} is based on the assumption that the spatial distribution of O$^+$ and H$^+$ are the same, when in reality, O$^+$ occupies a smaller volume of a typical \hii\ region than H$^+$ \citep[e.g.,][]{1992AJ....103.1330G}. In summary, recombination is unlikely to be a significant contributor to the observed \oi\ $\lambda 8449$ emission.

\subsection{Collisional excitation}

\begin{figure}
    \centering
    \includegraphics[width=8.5cm]{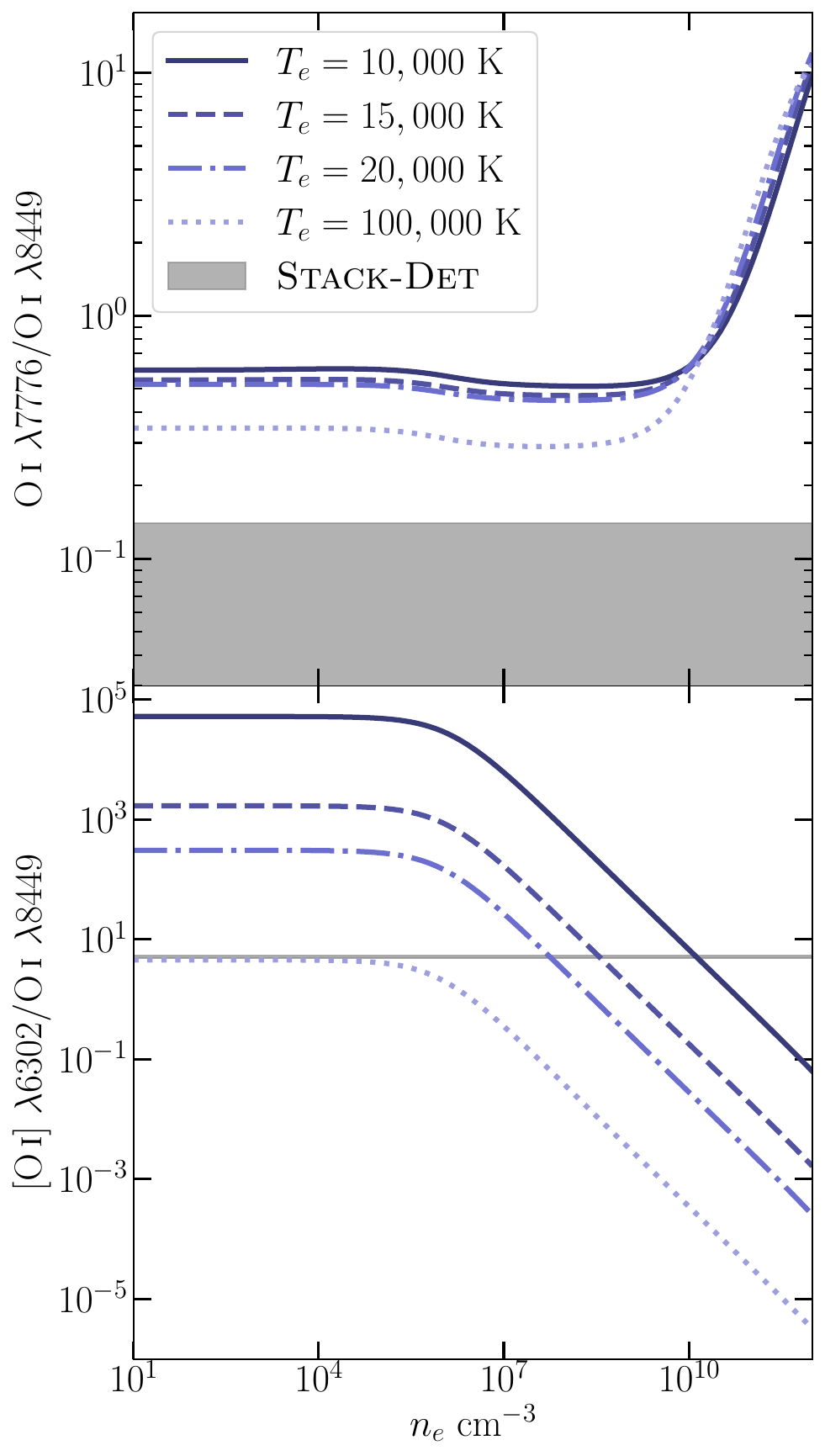}
    \caption{Oxygen line ratio predictions from {\sc PyNeb}. Measurements from the {\sc Stack-Det} composite are shaded in gray. ({\it Top}) Predicted \oi\ $\lambda 7776/$\oi\ $\lambda 8449$ ratio vs. $n_e$ and $T_e$. ({\it Bottom}) [\oi] $\lambda 6302$/\oi\ $\lambda 8449$ ratio vs. $n_e$ and $T_e$.}
    \label{fig:oi_temden}
\end{figure}

Another proposed excitation mechanism is the population of the 3p $\rm ^3P$ level by collisions between free electrons and O$^0$ atoms. \citet{1980ApJ...238...10G} explored signatures of collisionally excited \oi\ $\lambda 8449$, noting that for this mechanism, one would expect the ratio of \oi\ $\lambda 7776/$\oi\ $\lambda 8449$ to be $\sim$0.3, assuming an electron temperature of 10,000 K and adopting the collisional cross sections compiled by \citet{1977ApJ...214..785H}. \citet{1995ApJS...96..325B} carried out a more detailed analysis of the oxygen atom, computing effective collision strengths in a 13-level atomic model with 66 transitions, presenting line-strength predictions as a function of temperature and density. Utilizing the effective collision strengths from \citeauthor{1995ApJS...96..325B}, we calculate an expected ratio of \oi\ $\lambda 7776/$\oi\ $\lambda 8449=0.55\pm0.01$ using the {\sc PyNeb} package, adopting the $T_e$ and $n_e$ derived in Section \ref{sec:recombination}. We show the predicted \oi\ $\lambda 7776/$\oi\ $\lambda 8449$ ratio as a function of $T_e$ and $n_e$ in the top panel of Figure \ref{fig:oi_temden}. \oi\ $\lambda 7776$ is not detected in any object or stack from the AURORA sample. Five of the objects in this sample, as well as both stacks, yield upper limits on \oi\ $\lambda 7776/$\oi\ $\lambda 8449$ lower than 0.55. The {\sc Stack-NonDet} composite yields an upper limit of $<$0.54, which, formally, does not rule out collisional excitation for the non-detections. However, given that the upper limit on \oi\ $\lambda 7776/$\oi\ $\lambda 8449$ is $<$0.09 in the {\sc Stack-Det} composite, collisional excitation is unlikely to be the dominant contributor to \oi\ $\lambda 8449$ emission in the AURORA galaxies.

A more informative constraint on collisional excitation is obtained by interpreting the ratio of the forbidden [\oi] $\lambda 6302$ line to \oi\ $\lambda 8449$. Because the upper level of the [\oi] $\lambda 6302$ transition ($\rm 2p^4\ ^1D_2$) lies at a relatively low energy (1.967 eV), this line is easily collisionally excited in the outskirts of \hii\ regions, given the typical range of temperatures observed. If we assume that collisional excitation is responsible for 100\% of the [\oi] emission, and that [\oi] $\lambda 6302$ and \oi\ $\lambda 8449$ arise from the same gas, then the [\oi] $\lambda 6302$/\oi\ $\lambda 8449$ ratio serves as a valuable constraint on the collisional excitation of the O$^0$ atom.

Based on the $T_e$ and $n_e$ values derived in Section \ref{sec:recombination}, the inferred ratio of [\oi] $\lambda 6302$/\oi\ $\lambda 8449$ is $3200\pm1700$, adopting the collision strengths from \citet{1995ApJS...96..325B}. This value lies significantly above the ratios measured in all individual objects in the AURORA sample as well as the stack. To demonstrate the range of ISM conditions required to reproduce the observed [\oi] $\lambda 6302$/\oi\ $\lambda 8449$ in our sample, we show this ratio as a function of $T_e$ and $n_e$ in the bottom panel of Figure \ref{fig:oi_temden}. To reproduce the observed stacked [\oi] $\lambda 6302$/\oi\ $\lambda 8449$ ratio at typical ISM densities requires an electron temperature of 100,000 K, in extreme excess of plausible temperatures in even the hottest portions of \hii\ regions. To reproduce the [\oi] $\lambda 6302$/\oi\ $\lambda 8449$ ratio at more plausible \hii\ region temperatures, one must infer a density of $n_e\gtrsim10^8\ \rm cm^{-3}$, in excess of what is typically measured in \hii\ regions and PDRs \citep[e.g.,][]{1999RvMP...71..173H,2017MolAs...9....1W,2025MNRAS.541.1707T,2026arXiv260820339H}. Thus, to produce the \oi\ $\lambda 8449$ line with collisional excitation, one must assume implausibly extreme ISM conditions. We note that the [\oi] $\lambda 6302$ line can be enhanced by shocks, in addition to collisional excitation \citep[e.g.,][]{2017MNRAS.466.3217Z}. However, accounting for shock-enhanced [\oi] would only decrease the fraction of our measured [\oi] $\lambda 6302$/\oi\ $\lambda 8449$ ratio that is produced by collisional excitation, making the ratio more difficult to reproduce with collisions. Comparison of the [\oi] $\lambda 6302$/\oi\ $\lambda 8449 = 3200\pm1700$ ratio with our observed ratio from the stack suggests that collisional excitation contributes only $0.16\pm0.09$\% to the {\sc Stack-Det} \oi\ $\lambda 8449$ emission.

\subsection{Ly$\beta$ fluorescence}

The remaining processes, stellar continuum and Ly$\beta$ fluorescence, are expected to produce distinct patterns of \oi\ emission-line ratios. A key indicator to distinguish between these two mechanisms is the strength of the \oi\ $\lambda 11290$ line. In the case of Ly$\beta$ fluorescence, \oi\ $\lambda 11290$ is expected to have an equal photon flux to \oi\ $\lambda 8449$, since every transition from 3d $\rm ^3D^\circ$ to 3p $\rm ^3P$ (in the absence of other lines populating the 3p $\rm ^3P$ level) results in a subsequent emission of an \oi\ $\lambda 8449$ photon. This condition of equal photon flux thus sets a theoretical flux ratio of \oi\ $\lambda 11290$/\oi\ $\lambda 8449=0.749$. However, the observed ratio or upper limit of \oi\ $\lambda11290$/\oi\ $\lambda 8449$ in the stacks is significantly lower than the ratio predicted from Ly$\beta$ fluorescence. Furthermore, in individual objects, the 3 detected \oi\ $\lambda11290$/\oi\ $\lambda 8449$ ratios are significantly lower than the Ly$\beta$ fluorescence prediction, and 8 out of 10 upper limits are below this prediction as well.

A quantitative estimate of the contribution of Ly$\beta$ PAR to the emission of \oi\ $\lambda 8449$ can be found by following the analyses of \citet{1971MNRAS.153..393M} and \citet{1975ApJ...196..465G}. To explain the \oi\ $\lambda 8449$ emission in the Orion nebula, \citet{1975ApJ...196..465G} derived this expression for the predicted \oi\ $\lambda 8449$/H$\alpha$ ratio:\footnote{Note: \citet{1975ApJ...196..465G} originally provides the numerical values for this expression with respect to H$\beta$, having converted from H$\alpha$ using an unreported value for the Balmer decrement. We have re-expressed this equation here, assuming H$\alpha$/H$\beta=2.79$}

\begin{equation}
    \frac{F_{8449}}{F_{\rm H\alpha}} = 1.97\times 10^{-4} \left( \frac{\rm O/H}{6\times 10^{-4}} \right)
\end{equation}

This expression is obtained by integrating the volume emission coefficients for \oi\ $\lambda 8449$ and H$\alpha$ over an idealized model of Orion, and assuming that Ly$\beta$ PAR is the only excitation mechanism for \oi\ $\lambda 8449$. \citeauthor{1975ApJ...196..465G} also notes that the derivation of this ratio assumes that H$\alpha$ photons are only produced by the decay of Ly$\beta$ into H$\alpha$ and Ly$\alpha$, and thus the prediction should be interpreted as an upper limit. Adopting an oxygen abundance of $\rm O/H=(1.28\pm0.12)\times 10^{-4}$ for {\sc Stack-Det} based on the physical conditions calculated in Section \ref{sec:recombination}, we predict log(\oi\ $\lambda 8449$/H$\alpha)=-4.38\pm0.04$, which is similar to that derived for the case of recombination in Section \ref{sec:recombination}. This argument predicts that Ly$\beta$ fluorescence would only contribute $0.81\pm0.08\%$ to the observed \oi\ $\lambda 8449$ emission in {\sc Stack-Det}.

\citet{1980ApJ...238...10G} also calculated the predicted \oi\ $\lambda 8449$/H$\alpha$ ratio in the context of the broad-line region in AGN, yielding a value of log(\oi\ $\lambda 8449$/H$\alpha) = -2.82$, a much closer match to the values that we observe in AURORA. However, producing such strong \oi\ $\lambda 8449$ emission through Ly$\beta$ fluorescence requires extreme conditions, namely a very high optical depth to H$\alpha$ photons. As \citet{1980ApJ...238...10G} notes, Ly$\beta$ photons are converted to H$\alpha$ quickly after $\sim$10 scatterings, while $\sim$$5\times 10^4$ scatterings are required before Ly$\beta$ will encounter an oxygen atom and excite it to the 3d $\rm\ ^3D^\circ$ state. As a result, H$\alpha$ must be optically thick ($\tau_{\rm H\alpha} \sim 7\times 10^4$) to retain enough Ly$\beta$ photons within the nebula for the PAR process to be efficient. It is unlikely that this condition is satisfied in the majority of \hii\ regions such that the \oi\ $\lambda 8449$ line is measured at the strength we observe in both composite stacks comprising 58 star-forming galaxies as well as the 1602 galaxies from the DESI survey. As such, we rule out this mechanism as a significant contributor to \oi\ $\lambda 8449$ emission. With Ly$\beta$ fluorescence ruled out, we explore the remaining mechanism: stellar continuum fluorescence.

\subsection{Stellar continuum fluorescence}

Qualitatively, a proposed indicator of stellar continuum fluorescence of \oi\ $\lambda 8449$ is the presence of \oi\ $\lambda 13168$ emission that is stronger than $\lambda 11290$, but weaker than $\lambda 8449$, i.e., \oi\ $\lambda 8449>$ \oi\ $\lambda 13168 >$ \oi\ $\lambda 11290$ \citep[e.g.,][]{1975ApJ...196..465G,1991ApJ...368..468R}. This trend is borne out in the majority of the AURORA \oi\ sample as well as the stacks. For a quantitative comparison, we analyze the emission-line predictions performed by \citet{1975ApJ...196..465G} for an idealized model of the Orion nebula. For the case of continuum fluorescence, \citet{1975ApJ...196..465G} predicts an \oi\ $\lambda 8449$/H$\beta$ ratio of 0.009 which, assuming an intrinsic Balmer decrement of H$\alpha$/H$\beta=2.86$, yields an \oi\ strength of log(\oi\ $\lambda 8449$/H$\alpha)=-2.50$, consistent with the distribution of \oi\ $\lambda 8449$/H$\alpha$ values observed in the AURORA sample as well as CHAOS and DESI. \citet{1975ApJ...196..465G} also predicts a \oi\ $\lambda 13168$/\oi\ $\lambda 8449$ ratio of 0.33, and a \oi\ $\lambda 11290$/\oi\ $\lambda 8449$ ratio of 0.022 for Orion. The \oi\ $\lambda 13168$/\oi\ $\lambda 8449$ prediction is in good agreement with the {\sc Stack-Det} composite, and lower than, but the same order of magnitude as the {\sc Stack-NonDet} composite. The \oi\ $\lambda 11290$/\oi\ $\lambda 8449$ prediction is much smaller than the observed value in our stack as well as the three individual objects with detections, possibly suggestive of the need for a more sophisticated model to account for this line ratio. Despite the small differences between the \citet{1975ApJ...196..465G} predictions and our observations, this comparison suggests that stellar continuum fluorescence is the most compelling explanation for the \oi\ $\lambda 8449$ emission.

\subsection{Cloudy photoionization modeling}

To provide an independent set of predictions of the \oi\ line strengths, we model a mock \hii\ region and surrounding neutral ISM using {\sc Cloudy} \citep{2025RMxAA..61c.120G}. We utilize a spherical \hii-region model, fixing the gas-phase oxygen abundance to a value of $\rm 12+\log(O/H)=8.11$, as derived from the {\sc Stack-Det}-based physical conditions presented in Section \ref{sec:recombination}. Another input to the {\sc Cloudy} model is the ionization parameter, $\log(U)$. We derive this parameter from the {\sc Stack-Det} composite using the empirical relationship between $\log(U)$, $\rm 12+log(O/H)$, and dust-corrected [\oiii]\ $\lambda5008$/[\oii]\ $\lambda\lambda 3727,3730$ from \citet{2019ARA&A..57..511K}. For this calculation, based on the inferred $n_e$ and $T_e$, we assume $\log(P/k)=7.0$ and derive a value for the ionization parameter of $\log(U)=-2.72$. We fix the ionization parameter to this value in the {\sc Cloudy} simulation. We irradiate the cloud using the Binary Population and Stellar Synthesis \citep[BPASS;][]{2017PASA...34...58E,2018MNRAS.479...75S} v2.2.1 stellar population models, adopting a stellar age of 1 Myr, a stellar metallicity of $Z_*=0.001$, and an upper IMF cutoff of $300\ M_\odot$, including the effects of binary stars. We adopt an isobaric \hii-region model, varying the initial hydrogen density between $\log(\rm n_H/\rm cm^{-3})=[3,5]$. In this setup, the converged model typically produces a density profile that is roughly constant within the \hii\ region, equal to the initialized density value, and sharply increases at the \hii\ region boundary into the neutral zone. We thus refer to the initialized density as the \hii-region density. 

\begin{figure}
    \centering
    \includegraphics[width=8.6cm]{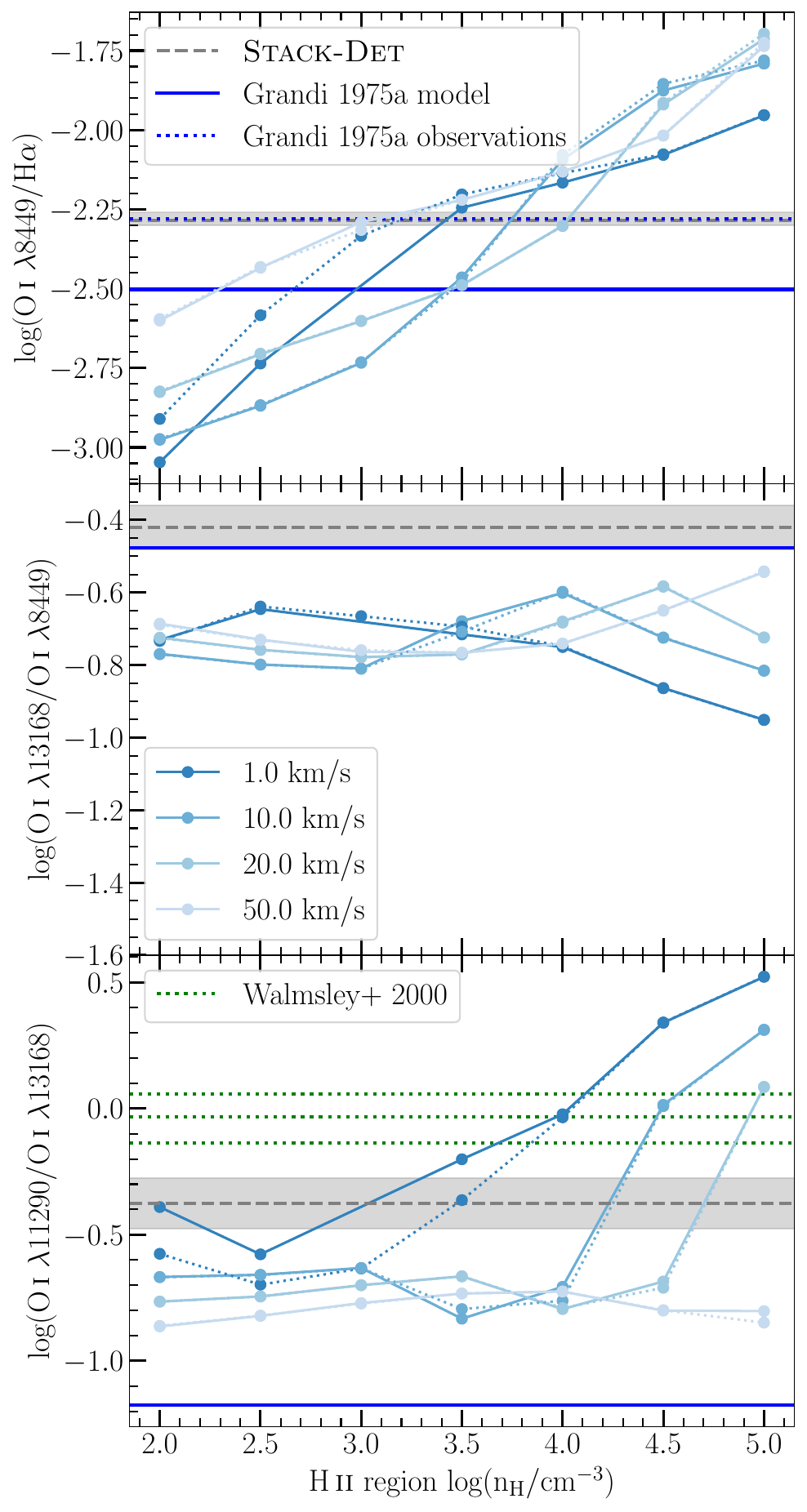}
    \caption{Cloudy model \oi\ emission-line ratios vs. initial \hii\ region density. The solid-line model curves show the {\sc Cloudy} runs with a cosmic-ray background included, while the dotted-line model curves show models that exclude cosmic rays. Measurements from the {\sc Stack-Det} spectrum are indicated by a gray dashed line, with the 1$\sigma$ confidence interval shaded in gray. For comparison, the \citet{1975ApJ...196..465G} observation and model predictions are shown as dotted and solid blue horizontal lines, respectively. Observations of the Orion bar from \citet{2000A&A...364..301W} are also shown for comparison as green dotted lines.}
    \label{fig:cloudy}
\end{figure}

We additionally vary the ISM turbulence, since the kinematics of the gas in the ISM modulates the absorption line widths, thereby impacting the amount of continuum energy that is absorbed and re-emitted in the fluorescent lines. We vary this property between [1,50] km/s. To sufficiently extend the simulation into the neutral ISM where the bulk of \oi\ emission originates, we set the cloud thickness to 2 pc, finding that this thickness sufficiently captures the extended \oi\ emission profile. The Galactic cosmic-ray background from \citet{2007ApJ...671.1736I} is included in the simulation. To allow for transitions from higher energy levels that feed into the 3p $^3$P level (e.g., from 4s $^3$S$^\circ$, 5s $^3$S$^\circ$, 6s $^3$S$^\circ$, 7s $^3$S$^\circ$, 4d $^3$D$^\circ$, 5d $^3$D$^\circ$, and 6d $^3$D$^\circ$, as in \citealt{1975ApJ...196..465G}), we include the lowest 100 energy levels of the O$^0$ atom in the simulation, with the 100th level corresponding to the 6g $\rm ^3G^\circ_4$ term. We show various resulting emission-line ratios in Figure \ref{fig:cloudy}.

In the top panel of Figure \ref{fig:cloudy}, we show the predicted \oi\ $\lambda 8449$/H$\alpha$ ratio as a function of \hii-region density and ISM turbulence. Apart from the lowest densities, the range of turbulence values that we explore yields line-ratio predictions that are comparable to each other within $\sim$0.3 dex. Our {\sc Cloudy} model predictions are largely consistent with the model prediction from \citet{1975ApJ...196..465G} for the Orion nebula at $n_e=4000\rm\ cm^{-3}$, shown as a solid dark blue line. Converting their reported measurement of \oi\ $\lambda 8449$/H$\beta$ to \oi\ $\lambda 8449$/H$\alpha$ assuming H$\alpha$/H$\beta=2.86$ (appropriate for a $T_e=10,000$ K nebula), we derive an \oi\ $\lambda 8449$/H$\alpha$ value for Orion (dark blue dotted line) that matches the {\sc Stack-Det} value almost exactly. These similarities suggest similar processes may be at work in Orion and in the AURORA star-forming galaxies. For \hii-region densities of $\gtrsim$10$^3\rm\ cm^{-3}$, the {\sc Cloudy} models predict log(\oi\ $\lambda 8449$/H$\alpha)> -2.75$, with higher densities yielding better agreement with the observed {\sc Stack-Det} value of $-2.28\pm0.02$. 

Informed by the constraints from the [\sii] ratio presented in Section \ref{sec:recombination}, adopting a value of $n_{\rm H}=10^{2.65}\rm\ cm^{-3}$ places {\sc Stack-Det} in the density regime where model predictions fall short of the {\sc Stack-Det} \oi\ $\lambda 8449$/H$\alpha$ ratio. However, observations have demonstrated that the density structure of the ISM can vary greatly between indicators, with high-ionization density indicators such as the C\thinspace{\sc iii}] doublet yielding higher densities by over an order of magnitude compared to [\sii] \citep[e.g.,][]{2025MNRAS.541.1707T}. Typically, in our simplified, isobaric \hii-region models, the density is roughly constant throughout the \hii\ region, only increasing significantly at the ionization front and into the surrounding neutral medium. Due to the non-trivial density structure of the ISM at $z\sim2$, a range of \hii-region densities can plausibly be considered for our simplistic model, and the \hii-region density does not necessarily have to match that derived from the [\sii] doublet. 

If we adopt a density of $n_{\rm H}\gtrsim10^{3.5}\rm\ cm^{-3}$, consistent with that adopted by \citet{1975ApJ...196..465G} for Orion, stellar continuum fluorescence can fully account for the observed \oi\ $\lambda 8449$/H$\alpha$ ratio in the {\sc Stack-Det} composite over a wide range of turbulence values. Given the uncertainties surrounding the effects of the \hii-region density structure on these results, future work employing more sophisticated \hii-region density profiles would prove as a valuable avenue to more robustly model \oi\ $\lambda 8449$ emission. As an additional exercise, we explore the effects of varying the nebular oxygen abundance, finding that this parameter has a minimal impact on the predicted \oi\ line ratios. The results of this exercise are shown in Appendix \ref{sec:appendix_a}. 

In addition to the \oi\ $\lambda 8449$/H$\alpha$ ratio, we show predictions for the \oi\ $\lambda 13168$/\oi\ $\lambda 8449$ ratio in the middle panel of Figure \ref{fig:cloudy}. The {\sc Cloudy} models shown in this Figure exhibit log(\oi\ $\lambda 13168$/\oi\ $\lambda 8449$) ratios that are smaller than the {\sc Stack-Det} value of $-0.42\pm0.06$, differing by $\sim 0.1-0.5$ dex across a wide range of hydrogen densities and turbulence values. The \citet{1975ApJ...196..465G} model prediction of log(\oi\ $\lambda 13168$/\oi\ $\lambda 8449)=-0.48$ is closer to the observed {\sc Stack-Det} ratio, differing by only $0.06\pm0.06$ dex. We also examine the \oi\ $\lambda11290$/\oi\ $\lambda 13168$ ratio, shown in the bottom panel of Figure \ref{fig:cloudy}. In the case of stellar continuum fluorescence, this ratio is predicted to be less than unity \citep{1975ApJ...196..465G,1991ApJ...368..468R}. Indeed, \oi\ $\lambda 11290$/\oi\ $\lambda 13168 < 1$ in the {\sc Stack-Det} composite. For turbulence values $>$1 km/s, {\sc Cloudy} underpredicts the \oi\ $\lambda 11290$/\oi\ $\lambda 13168$ ratio, differing by $\sim$0.2--0.5 dex 
below the {\sc Stack-Det} ratio at $\rm n_H \lesssim 10^4\ cm^{-3}$. The discrepancy between {\sc Stack-Det} and the ratio predicted by \citet{1975ApJ...196..465G} is even larger, differing by $0.80\pm0.10$ dex.

The \oi\ $\lambda 11290$/\oi\ $\lambda 13168$ ratio can be explained in the case of an \hii\ region with $n_{\rm H}=10^2\rm\ cm^{-3}$ and 1 km/s of turbulence. However, the predicted \oi\ $\lambda 8449$/H$\alpha$ ratios disfavor such a low \hii-region density. Adopting densities of $n_{\rm H}\gtrsim10^{4}\rm\ cm^{-3}$ and $<$50 km/s of turbulence also yields \oi\ $\lambda 11290$/\oi\ $\lambda 13168$ ratios that are close to the observed value. As such, the \oi\ $\lambda 11290$/\oi\ $\lambda 13168$ ratio suggests a density of $\sim$$10^4-10^5\rm\ cm^{-3}$, roughly consistent, though slightly higher than the densities one would infer based on the \oi\ $\lambda 8449$/H$\alpha$ ratio alone.

\citet{2000A&A...364..301W} measured values of log(\oi\ $\lambda 11290$/\oi\ $\lambda 13168)=-0.14$ to $0.06$ for three pointings in the Orion Bar, exceeding the ratio predicted by \citet{1975ApJ...196..465G} by an order of magnitude. For most combinations of turbulence and \hii-region density, the {\sc Cloudy} models also predict a lower \oi\ $\lambda 11290$/\oi\ $\lambda 13168$ ratio than observed by \citet{2000A&A...364..301W}. The {\sc Cloudy} models that best agree with these observations are characterized by either a low turbulence ($\sim$1 km/s), or a high density at fixed turbulence. It is plausible to consider the possibility that the \oi\ $\lambda 11290$/\oi\ $\lambda 13168$ ratio is higher in the Orion bar than in the $z\sim 2$ AURORA sample because the ISM at $z\sim 2$ is typically more turbulent than at $z\sim 0$ \citep{2020ARA&A..58..661F}. A valuable improvement on the {\sc Cloudy} models would be to implement more realistic density and turbulence gradients in the simulated nebula, since both of these quantities can strongly influence \oi\ $\lambda 11290$/\oi\ $\lambda 13168$, and should accordingly be modeled in more careful detail. Future work focusing on such improvements will be necessary in order to fully understand \oi\ production in star-forming galaxies.

In summary, comparing observed \oi\ line ratios in AURORA and Orion with {\sc Cloudy} models suggests that \hii-region densities in the range $\sim$$10^{3.5}-10^5\rm\ cm^{-3}$ are necessary to explain the observations. Reproducing the observed \oi\ $\lambda 13168$/\oi\ $\lambda 8449$ ratio remains a challenge, since no {\sc Cloudy} models presented in this work match the observed ratio. This discrepancy may indicate the presence of additional physical considerations that are not currently captured by the models. However, compared to the other excitation mechanisms (i.e., recombination, collisional excitation, and Ly$\beta$ PAR), stellar continuum fluorescence is the most compelling in its ability to predict the observed \oi\ line strengths.

\section{Discussion}\label{sec:discussion}

Stellar continuum fluorescence as a production mechanism for \oi\ $\lambda 8449$ in the AURORA sample of star-forming galaxies has several interesting astrophysical implications. Firstly, we have ruled out Ly$\beta$ fluorescence as a dominant mechanism producing \oi\ $\lambda 8449$ based on the observed strength of \oi\ $\lambda 13168$ relative to \oi\ $\lambda 11290$, as well as the observed \oi\ $\lambda 8449$/H$\alpha$ ratio, which diverges significantly from predictions for Ly$\beta$ fluorescence in an \hii\ region presented by \citet{1975ApJ...196..465G}. As such, though \oi\ $\lambda 8449$ is commonly observed arising from the broad-line regions of AGN due to Ly$\beta$ fluorescence and collisional excitation \citep[e.g.,][]{1980ApJ...238...10G,1989ApJ...342..235R,2002ApJ...572...94R}, its presence in the spectra of star-forming galaxies is not an indicator of the aforementioned processes. 

Notably, we observe in our sample a strong correlation between the \oi\ $\lambda 8449$ and H$\alpha$ line strengths, evidenced by the relatively narrow range of observed \oi\ $\lambda 8449$/H$\alpha$ ratios shown in Figure \ref{fig:oi_ha}. This tight correlation between \oi\ and H$\alpha$ emission can indeed arise due to Ly$\beta$ fluorescence, as suggested by \citet{2025arXiv251121820D} in the case of a large sample of LRDs. However, given that we have strongly ruled out contributions from Ly$\beta$ fluorescence in favor of stellar continuum fluorescence, we highlight the necessity of constraining the \oi\ $\lambda 8449$ production mechanism using both the \oi\ $\lambda 11290$ and \oi\ $\lambda 13168$ emission lines, given that a tight \oi\ $\lambda 8449$-H$\alpha$ correlation is not itself an indicator of Ly$\beta$ fluorescence.

Additionally, the detection of a {\it permitted} oxygen line in star-forming galaxies may initially lead to the assumption that this line comes from recombination. In this case, the \oi\ $\lambda 8449$/H$\alpha$ ratio would serve as a valuable metallicity indicator, since recombination-line metallicities have a weaker temperature dependence than those derived from the ``direct" $T_e$-based method \citep[e.g.,][]{2019A&ARv..27....3M}, and are less susceptible to the effects of temperature fluctuations that would bias derived abundances \citep[e.g.,][]{1967ApJ...150..825P,2003MNRAS.338..687T}. However, the non-detection of several other recombination lines such as \oi\ $\lambda 7776$, as well as the extremely low theoretical recombination-based \oi\ $\lambda 8449$/H$\alpha$ ratio derived in Section \ref{sec:recombination} securely rule out recombination as the dominant process contributing to \oi. Thus, \oi\ $\lambda 8449$/H$\alpha$ may be of limited use as a metallicity indicator for star-forming galaxies, and will certainly lead to unreasonable estimates if recombination is assumed to be the dominant mechanism \citep{2023ApJ...958L..11S}.

Beyond the local Universe, the \oi\ $\lambda 8449$ feature has now been observed with JWST in a variety of sources. In the Sunburst Arc at $z=2.37$, \citet{2025A&A...698A..16C} measured this line in a highly magnified source referred to as ``Godzilla," noting a highly elevated \oi\ $\lambda 8449$/H$\beta$ ratio of $20.75\pm 0.15\%$. This ratio is much larger than the values that we find to be characteristic of our sample of star-forming galaxies. Indeed, \citet{2025A&A...698A..16C} suggest that the mechanism powering \oi\ $\lambda 8449$ in Godzilla is Ly$\beta$ fluorescence, resembling closely the processes occurring in the Weigelt blobs in the luminous blue variable $\eta$ Car \citep{2005MNRAS.364..731J}. \citet{2020MNRAS.499L..67V} also find evidence of PAR due to Ly$\alpha$ in this object, detecting the emission of several pumped Fe lines. Thus, the production mechanism for \oi\ $\lambda 8449$ in Godzilla differs from what we observe in our sample, likely due to the extreme conditions (i.e., $\gtrsim$$10^6\rm\ cm^{-3}$) found in this source. Because this emission line is now more routinely observed in high-redshift sources thanks to the exquisite sensitivity of JWST, it is critical to understand its production mechanism in each class of object, given that the different \oi\ production mechanisms are indicators of vastly different physical conditions.

We note that in the analysis of \citet{2025A&A...698A..16C}, they also present rest-UV observations covering the \oi\ $\lambda1302,\lambda1305,\lambda1306$ triplet, observed in both emission and absorption. Though we measure the \oi\ emission to be weaker in typical star-forming galaxies than in Godzilla (as quantified by the \oi\ $\lambda 8449$/H$\alpha$ ratio), the complex \oi\ and Si\thinspace{\sc ii} profiles presented by \citet{2025A&A...698A..16C} may highlight the general need for a more careful treatment of both emission and absorption components when fitting this feature in down-the-barrel observations of $z\sim 2$ galaxies \citep[e.g.,][]{2010ApJ...717..289S, 2021ApJ...920...95D,2022ApJ...926...31R}. Overall, this work highlights the valuable constraints on the production mechanism of \oi\ $\lambda 8449$ obtained from measuring the \oi\ $\lambda 11290$ and \oi\ $\lambda 13168$ lines. With these constraints, it is possible to distinguish between a variety of distinct physical conditions that all produce the \oi\ $\lambda 8449$ emission line.

\section{Conclusions}

We present detections of the permitted \oi\ $\lambda 8449$ line in 14 star-forming galaxies, as well as stacks comprising 58 galaxies from the AURORA survey. We explore evidence for four candidate excitation mechanisms: recombination, collisional excitation, Ly$\beta$ fluorescence, and stellar continuum fluorescence. We determine that stellar continuum fluorescence provides the most compelling explanation for the production of this emission line. Our primary conclusions are as follows:

\begin{enumerate}
    \item The \oi\ $\lambda 8449$ line is present in the stack of all galaxies in which the line was not individually detected ({\sc Stack-NonDet}). As such, it is likely that \oi\ $\lambda 8449$ emission (at $\sim$0.5\% the strength of H$\alpha$)  is a ubiquitous feature of star-forming galaxies at $z\sim 2$. Additionally, the detection of this line in the more nearby Universe ($z<0.1619$) in 1602 galaxies from DESI DR1 as well as local \hii\ regions from the CHAOS survey suggests that this fluorescent line is a common feature of star-forming galaxies across cosmic time.
    \item We determine that the dominant production mechanism for \oi\ $\lambda 8449$ in star-forming galaxies is stellar continuum fluorescence, as evidenced by the detection of \oi\ $\lambda 13168$. We estimate that recombination, collisional excitation, and Ly$\beta$ fluorescence contribute $1.0\%\pm0.2\%$, $0.16\%\pm0.09\%$, and $0.81\% \pm 0.08\%$ to the \oi\ $\lambda 8449$ emission, respectively.
    \item Simultaneously modeling \oi\ $\lambda 8449$, \oi\ $\lambda 11290$, and \oi\ $\lambda 13168$ is challenging when using the isobaric {\sc Cloudy} models that we present in this analysis. Improving upon these simplified models with a more complex density profile and placing tighter constraints on ISM turbulence may serve as valuable avenues for improving predictions of the relative strengths of these three lines.
\end{enumerate}

JWST observations continue to enable the characterization of the $z\sim 2$ star-forming galaxy population in unprecedented detail through the measurement of faint spectroscopic features. Characterizing and accurately interpreting the origins of these features is critical for building a comprehensive picture of the conditions in the ISM. In the case of \oi\ $\lambda 8449$, because this feature is seen in a variety of sources and can be produced by very different conditions, constraining the production mechanism for this line in star-forming galaxies is valuable for ruling out the extreme and unusual conditions that would be required for processes such as Ly$\beta$ PAR to contribute. Additionally, because faint metal recombination lines are often used to derive metallicities with little dependence on density or temperature, accurately distinguishing between recombination and contributions from fluorescence is critical for deriving robust metal abundances \citep[e.g.,][]{2003MNRAS.338..687T}. In summary, though future work remains to be done to better reproduce the observed \oi\ line emission strengths in the AURORA sample, stellar continuum fluorescence offers the most compelling explanation for the presence of this line in star-forming galaxies.

\begin{acknowledgments}
We thank Noah Rogers for kindly sharing the full CHAOS emission-line catalog. We also thank Mirko Curti and Anna de Graaff for useful discussions surrounding this work. We also acknowledge support from NASA grants JWST-GO-01914 and JWST-GO-03833, and NSF AAG grants 2009313, 2009085, 2307622, and 2307623. This material is based upon work supported by the National Science Foundation under Award No. 2602699 for author AJP. This work is based on observations made with the NASA/ESA/CSA James Webb Space Telescope as well as the NASA/ESA Hubble Space Telescope. The data were obtained from the Mikulski Archive for Space Telescopes at the Space Telescope Science Institute, which is operated by the Association of Universities for Research in Astronomy, Inc., under NASA contract NAS5-03127 for JWST and NAS 5–26555 for HST. The specific observations analyzed can be accessed via \dataset[doi:10.17909/evwk-m381]{https://archive.stsci.edu/doi/resolve/resolve.html?doi=10.17909/evwk-m381}. This work used computational and storage services associated with the Hoffman2 Cluster, which is operated by the UCLA Office of Advanced Research Computing’s Research Technology Group. This research used data obtained with the Dark Energy Spectroscopic Instrument (DESI). DESI construction and operations is managed by the Lawrence Berkeley National Laboratory. This material is based upon work supported by the U.S. Department of Energy, Office of Science, Office of High-Energy Physics, under Contract No. DE–AC02–05CH11231, and by the National Energy Research Scientific Computing Center, a DOE Office of Science User Facility under the same contract. Additional support for DESI was provided by the U.S. National Science Foundation (NSF), Division of Astronomical Sciences under Contract No. AST-0950945 to the NSF’s National Optical-Infrared Astronomy Research Laboratory; the Science and Technology Facilities Council of the United Kingdom; the Gordon and Betty Moore Foundation; the Heising-Simons Foundation; the French Alternative Energies and Atomic Energy Commission (CEA); the National Council of Humanities, Science and Technology of Mexico (CONAHCYT); the Ministry of Science and Innovation of Spain (MICINN), and by the DESI Member Institutions: \url{www.desi.lbl.gov/collaborating-institutions}. The DESI collaboration is honored to be permitted to conduct scientific research on I'oligam Du'ag (Kitt Peak), a mountain with particular significance to the Tohono O'odham Nation. Any opinions, findings, and conclusions or recommendations expressed in this material are those of the author(s) and do not necessarily reflect the views of the U.S. National Science Foundation, the U.S. Department of Energy, or any of the listed funding agencies.
\end{acknowledgments}




%
\facilities{JWST(NIRSpec, NIRCam), HST(WFC3, ACS)}

\software{astropy \citep{2013A&A...558A..33A,2018AJ....156..123A,2022ApJ...935..167A},  
          Cloudy \citep{2013RMxAA..49..137F}, 
          {\sc PyNeb} \citep{2015A&A...573A..42L}
          }


\appendix

\section{Varying Cloudy Oxygen abundances}\label{sec:appendix_a}

\begin{figure*}
    \centering
    \includegraphics[width=\linewidth]{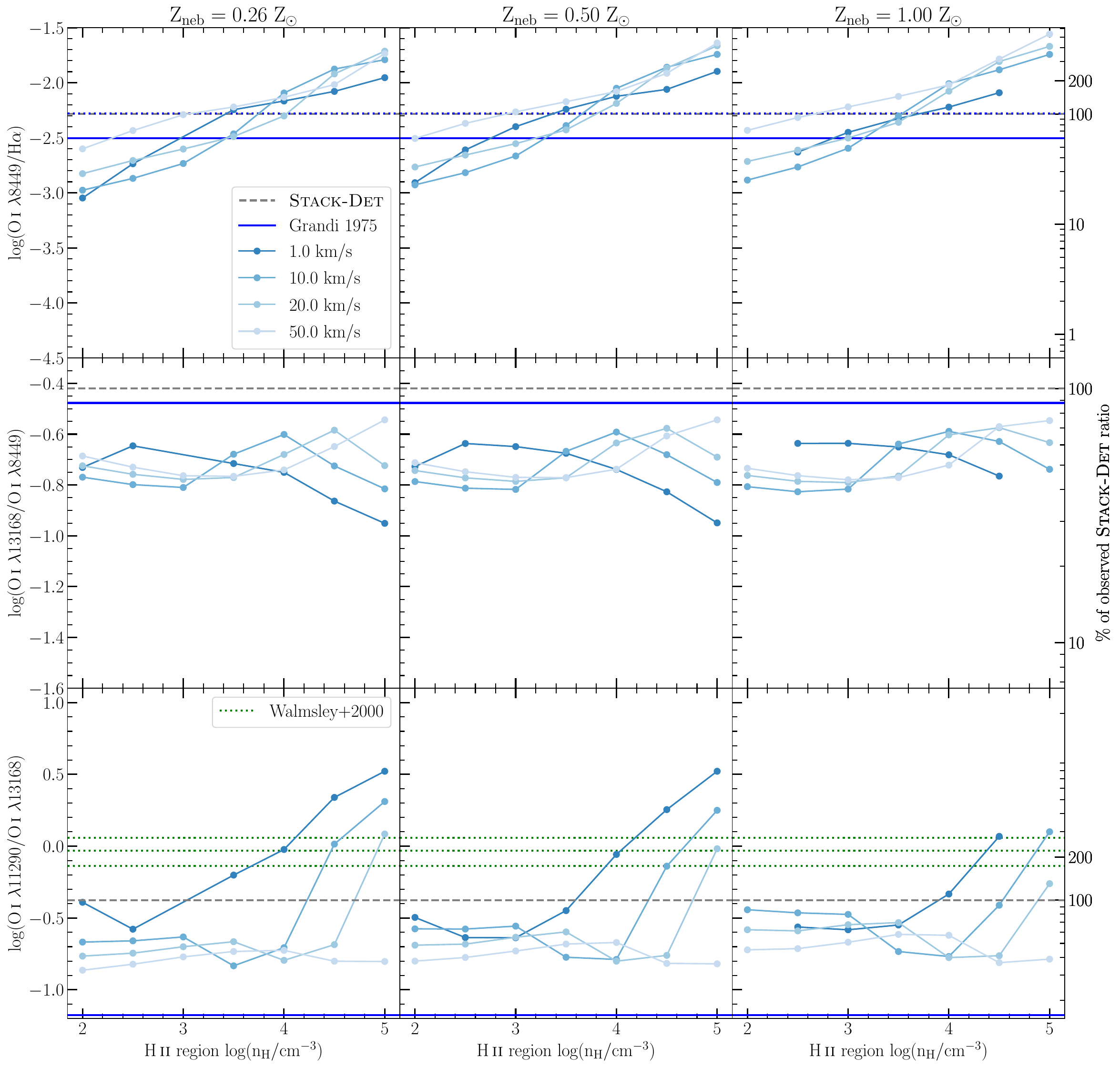}
    \caption{{\sc Cloudy} models adopting various nebular oxygen abundances. The $\rm Z_{neb}=0.26\ Z_\odot$ column is the fiducial set of models presented in the main text.}
    \label{fig:cloudy_appendix}
\end{figure*}

In Figure \ref{fig:cloudy_appendix}, we show the effects of varying the nebular oxygen abundance on the various \oi\ line ratios that we consider. The subplots in the left column show the same model curves as those presented in Figure \ref{fig:cloudy}. We additionally consider oxygen abundances of 0.5 $\rm Z_\odot$ and 1.0 $\rm Z_\odot$. Overall, varying the nebular metallicity over this specified range does not have a significant impact on the predicted \oi\ $\lambda 8449$/H$\alpha$ ratios. In principle, a Monte Carlo radiative transfer calculation would be required to robustly capture the line fluorescence channels. However, we present the {\sc Cloudy} model predictions at different oxygen abundances as a rough estimate of the effect of varying metallicity.

\bibliography{sample701}{}
\bibliographystyle{aasjournalv7}



\end{document}